\documentclass[twocolumn,prb,reprint,showpacs,preprintnumbers,amsmath,amssymb,superscriptaddress]{revtex4-1}
\usepackage{graphicx}
\usepackage{dcolumn}
\usepackage{bm}
\usepackage{amsmath}
\let\oldAA\AA
\renewcommand{\AA}{\text{\normalfont\oldAA}}

\usepackage{amsmath}
\usepackage{subfigure}
\usepackage{hyperref}
\usepackage[normalem]{ulem}
\hypersetup{
    colorlinks,
    citecolor=blue,
    filecolor=black,
    linkcolor=blue,
    urlcolor=blue
}
\usepackage{epstopdf}
 \usepackage{relsize}
\usepackage{ragged2e}

\newcommand*{\rom}[1]{\expandafter\@slowromancap\romannumeral #1@}
\begin{document}


\title{Incoherent scattering can favorably influence energy filtering in nanostructured thermoelectrics.}

\author{Aniket Singha}

\affiliation{%
Department of Electrical Engineering,\\
Indian Institute of Technology Bombay, Powai, Mumbai-400076, India\\
}%


\author{Bhaskaran Muralidharan}%
\email{bm@ee.iitb.ac.in}
\affiliation{%
Department of Electrical Engineering,\\
Indian Institute of Technology Bombay, Powai, Mumbai-400076, India\\
}%



\date{\today}

\begin{abstract}
Investigating in detail the physics of energy filtering through a single planar energy barrier  in nanostructured thermoelectric generators, we reinforce the non-trivial result that the anticipated enhancement in  generated power at a given  efficiency via energy filtering is a characteristic of systems dominated by incoherent scattering and is absent in ballistic devices. In such cases, 
 assuming an energy dependent relaxation time $\tau(E)=kE^r$, we show that there exists a minimum value $r_{min}$ beyond which  generation can be enhanced by embedding nanobarriers. For bulk generators with embedded nanobarriers, we delve into the details of inter sub-band scattering and show that it has finite contribution to the enhancement in  generation. We subsequently discuss the realistic aspects, such as the effect of  smooth transmission cut-off  and  show that for $r>r_{min}$, the optimized energy barrier is just {\it{sufficiently wide enough}} to scatter off low energy electrons, a very wide barrier being detrimental to the  performance.  Analysis of the obtained results should provide general design guidelines for  enhancement in thermoelectric  generation via energy filtering.  Our non-equilibrium approach is typically valid in the absence of local quasi-equilibrium and hence sets the stage for future advancements in thermoelectric device analysis, for example,  Peltier cooling near a barrier interface.   
\end{abstract}

\flushbottom
\maketitle
%
%
\thispagestyle{empty}

\section*{Introduction}

\begin{figure*}[!htb]
\centering
\includegraphics[scale=1.3]{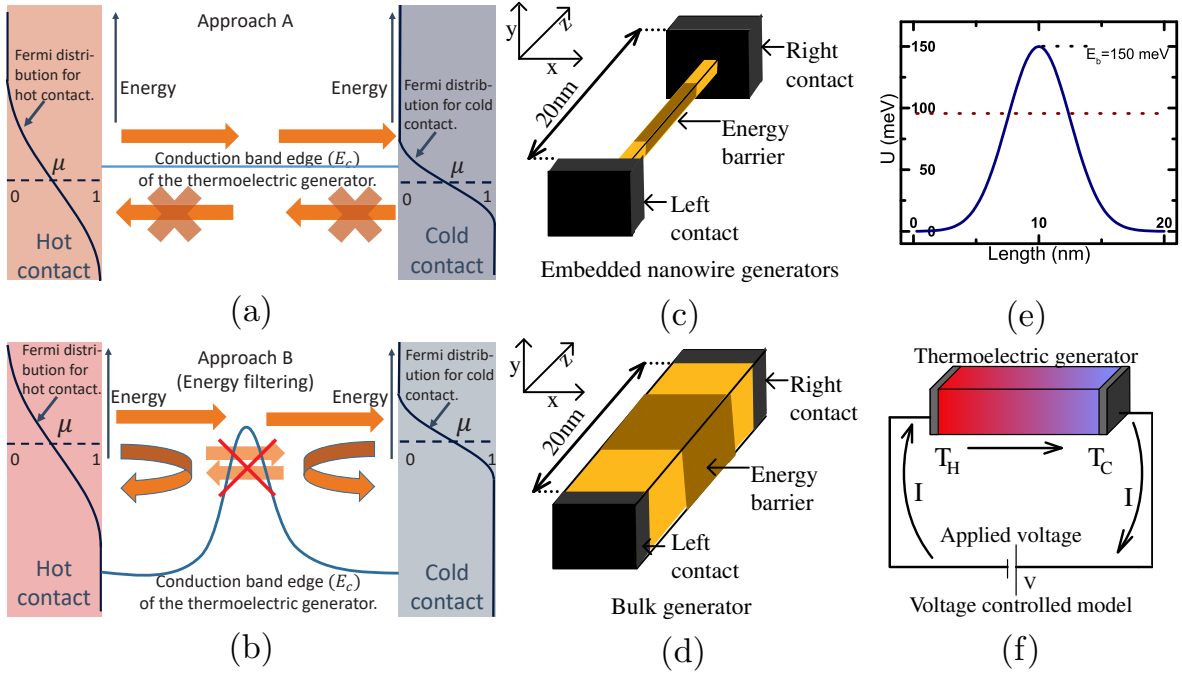}

\caption{(a-b) Schematics depicting the two common approaches of improving the thermoelectric performance. (a) Approach A: Enhancing the thermoelectric generation by tuning the position of Fermi energy near the conduction band edge and  (b) Approach B: enhancing the thermoelectric performance by  energy filtering via embedding a nano barrier within the thermoelectric generator. The hot and cold contacts are assumed to be macroscopic bodies in equilibrium with the quasi Fermi energy $\mu$ at temperature $T_H$ and $T_C$ respectively. (c-d) The device used for simulation, with a device region of length  $20nm$ comprising (c) embedded nanowire thermoelectric generator (d) bulk thermoelectric generator. The shaded region in both the cases represents the embedded barrier (in case of  Approach B). (e) The band profile of the device region of length $20nm$, embedded with a  Gaussian energy barrier of height $E_b=150meV$ and width $\sigma_w=2.7nm$. The brown dotted line shows the Fermi energy of the device for the case $E_b-\mu_0=2k_BT$. (f) Schematic of the voltage controlled model used to simulate the power-efficiency trade-off points.  }
\label{fig:schem_filtering}
\end{figure*}

\indent An important direction in the context of electronic engineering to enhance the performance of nanostructured thermoelectric generators \cite{hicks1,hicks2,dresselhaus,goldsmid,rev_sofo_mahan,poudel,snyder,heremans,Mona_Z,zimb}, is to utilize the physics of electronic energy filtering through nanoscale barriers and nanoinclusions \cite{neophytou_inelastic,nanoinclusion1,nanoinclusion2,nanoinclusion3,nanoinclusion4,nanoinclusion5,nanoinclusion6,kim1,kim2,theory1,theory2,theory3,barrier1,barrier2}. To put it simply, energy filtering aims to provide a unidirectional flow of electrons from the hot contact to the cold contact while prohibiting the reverse flow of electrons, which occurs typically when lower energy electrons are scattered off due to the interface potentials \cite{theory1,theory2,theory3,Mona_Z,vashaee}. 
\indent In the case of semiconductors,  there is the flexibility of varying the equilibrium Fermi energy via appropriate doping. In such a case, a `good thermoelectric' as schematized in Fig. \ref{fig:schem_filtering}(a) is ideally achieved by tuning the Fermi energy near the conduction band edge so that the resulting transport is devoid of electrons below the Fermi energy. This off-resonant conduction typifies good thermoelectric behavior and has been the object of several initial proposals \cite{hicks1,hicks2,dresselhaus,rev_sofo_mahan,heremans}.   Such an approach however leads to a drastic reduction in the conductivity while enhancing the Seebeck coefficient. We will refer this as \emph{Approach A}. Energy filtering, as schematized in Fig. \ref{fig:schem_filtering}(b), on the other hand, strives to achieve a desirable performance via engineering nano-barriers
\cite{nanoinclusion1,nanoinclusion4,nanoinclusion5,kim1,kim2,rev1,rev2,rev3,rev4,theory1,theory2,theory3,barrier1}. In this case, the Fermi level resides inside the conduction band and the principal aim is to enhance the Seebeck coefficient by reflecting low energy electrons from the energy barrier. This approach also promotes a unidirectional flow of electrons from the hot contact to the cold contact. We will refer to this as \emph{Approach B}. Ideally, we can say that energy filtering is successful if the latter yields a better performance than the former, especially with a thinner barrier. \\
\indent  While recent works throw some light on this topic \cite{kim1,kim2,rev1,rev2,rev3,rev4,theory1,theory2,theory3,barrier1,barrier2}, we believe that a few aspects about energy filtering require attention: 1) Most of the current work is based on a linear response analysis of the Seebeck coefficient despite the fact that the regions in the vicinity of the barrier are strongly out of equilibrium. Linear response analysis typically masks the crucial transport physics that determines the delivered terminal power output and efficiency of the generator \cite{leij,sothmann,bm,nakpathomkun,jordan2,mypaper,akshay}. 2) The role of various scattering mechanisms contributing to the physics of energy filtering is still unclear. 3) A generalized picture of the physics of energy filtering independent of various material parameters is also unclear. In this paper, our focus is hence to develop a general and intuitive understanding of energy filtering and to systematically point out the role played by various scattering mechanisms. Besides that our work uses the non-equilibrium Green's function method which accounts for the non-equilibrium nature of transport and directly evaluates the power and efficiency to provide an overall picture of the device operation. \color{black} \\
\indent Our goal here is to present important clarifications on the aforesaid aspects. As a principal contribution, we clarify that the power generation enhancement via energy filtering in both nanowire and bulk devices is dependent on a specific property of the scattering mechanism, which we call \emph{the order of scattering}. It is first shown that for the ballistic case,  both Approach A and Approach B lead to an identical performance and hence energy filtering is of limited use. On the other hand, in the diffusive limit for both nanowire and bulk thermoelectric generators, it is shown that the type of scattering mechanism is the principal deciding factor in order to gauge the advantage gained via Approach B.  Assuming an energy dependent relaxation time $\tau(E)=kE^r$,  we then show that there is a minimum value of the exponent $r$ (termed $r_{min}$),  beyond which energy filtering via Approach B  leads to a better enhancement in the generated power compared to Approach A. For such cases, the generated power at a given efficiency increases with an increase in the height of the embedded energy barrier. In addition, we show that for bulk thermoelectric generators with embedded energy barriers, electronic scattering between the high energy and low energy sub-bands enhances the generated power. We also discuss some practical aspects, like the adverse effects of a smooth transmission cut-off due to finite barrier width and investigate further to conclude   in case of $r>r_{min}$, the thermoelectric performance peaks when the energy barrier is made {\it{ just sufficiently wide enough}} to scatter off the low energy electrons.  \\
\indent  We utilize the non-equilibrium Green's function (NEGF) formalism to deduce various currents, following which a direct calculation of power and efficiency is performed using the following equations:
\begin{equation}
P=I_C \times V
\label{eq:pow}
\end{equation} 
\begin{equation}
\eta=\frac{P}{I_{Qe}},
\label{eq:eff}
\end{equation}
where $I_C$ is the charge current, $I_{Qe}$ is the electronic heat current at the hot contact and $V$ is the applied voltage assuming a voltage controlled set up described in recent literature\cite{leij,sothmann,bm,nakpathomkun,jordan2,mypaper,akshay}.   Since thermal conductivity due to phonon doesn't vary significantly with the method employed to improve the thermoelectric performance (that is, Approach A or Approach B), we have simplified our calculations in \eqref{eq:eff} by neglecting the degradation in efficiency due to phonon heat conductivity. 
We consider thermoelectric generators in which the active regions are smaller than the energy relaxation lengths \cite{kim1} such that the energy current is almost constant throughout the device region.  This assumption simplifies the discussion to a great extent and aids in understanding the physics of energy filtering from an intuitive mathematical viewpoint. \color{black}  For the purpose of the simulations, we use the band parameters of the $\Delta_2$ valley of lightly doped silicon \cite{book1} with  a longitudinal effective mass, $m_l=m_e$, and a transverse effective mass, $m_t=0.2m_e$, $m_e$ being the mass of a free electron. The transverse geometries of the device region considered here include bulk, where the transverse extent is infinite and nanowires, where the transverse extent consists of only one sub-band. A schematic of the generic device structure used  is shown in Fig. \ref{fig:schem_filtering}~(c) for nanowire generators and Fig. \ref{fig:schem_filtering}~(d) for bulk generators.    The band diagram schematic of our energy filtering based thermoelectric generator is shown in Fig. \ref{fig:schem_filtering} (e). It consists of a $20nm$ long doped semiconductor thermoelectric generator  with an embedded Gaussian energy barrier (U). 
\[
U=E_b exp\left(-\frac{(z-z_0)^2}{2\sigma_w^2}\right),
\]
where $z_0=L/2$ is the the mid-point of the device region, $L$ being the total length of the thermoelectric generator. The  leads or contacts connected to the $20nm$ long device in Fig \ref{fig:schem_filtering} (c) and (d) are assumed to be reflection-less macroscopic bodies of infinite cross-section, in equilibrium with their  respective lattice temperatures and electrochemical potentials (quasi-Fermi levels). For simulation, we use a voltage controlled model, shown in Fig. \ref{fig:schem_filtering} (f), where by varying the bias voltage continuously a the current flow is emulated, to generate a set of points in the the power-efficiency ($\eta$)  plane \cite{nakpathomkun,whitney} for a particular position of the equilibrium electrochemical potential (or equivalently the Fermi energy) $\mu_0$. \\
\indent In order to assess the relative efficacy of power generation due to energy filtering and compare the relative benefit of Approach B compared to Approach A, we now define a metric called the \emph{filtering coefficient} ($\lambda$) defined as 
\begin{equation}
\lambda(\eta)=\frac{P_B(\eta)}{P_A(\eta)},
\label{eq:fc}
\end{equation}
where $P_A(\eta)$ and $P_B(\eta)$ are the maximum power densities obtained at efficiency $\eta$ via Approach A and Approach B  respectively, while $P_A(\eta)$ and $P_B(\eta)$ are taken along the operating line of our device \cite{mypaper}.  It should be noted that although we  use some specific parameters for simulation, the qualitative discussion  as well as the trends noted in the simulated results  are general and are valid irrespective of material specific parameters  for a particular scattering mechanism. \color{black} \\

\section*{Methods}
\indent In order to perform the transport calculations to be presented, we employ the NEGF transport formalism under the self-consistent Born approximation \cite{dattabook,Lake_Datta} to incorporate scattering in the device region (details given in the supplementary material). We employ the single particle Green's function $G(\overrightarrow{k_{m}},E)$, for each transverse sub-band $m$ \cite{dattabook}, evaluated from the device Hamiltonian matrix [H] given by:
\begin{gather}\label{eq:negf_main}   
G(\overrightarrow{k_{m}},E)=[EI-H-U-E_m-\Sigma(\overrightarrow{k_{m}},E)]^{-1} ,\nonumber \\
\Sigma(\overrightarrow{k_{m}},E)=\Sigma_L(\overrightarrow{k_{m}},E)+\Sigma_R(\overrightarrow{k_{m}},E)+\Sigma_s(\overrightarrow{k_{m}},E), 
\end{gather}
where $[H]=[H_0]+U$, with $[H_0]$ being the device tight-binding Hamiltonian matrix constructed using effective mass approach \cite{dattabook} and $I$ being the identity matrix of the same dimension as the Hamiltonian.  The free variable denoting the energy of electronic wavefunction is $E$ and the spatial variation in the conduction band minimum is described by the matrix $U$.  The sub-band energy  of the $m^{th}$ sub-band is $E_m=\frac{\hslash^2k_m^2}{2m_t}$. The vector $\overrightarrow{k_{m}}$ represents the wavevector of the electron in the transverse direction for the $m^{th}$ mode. The net scattering self energy matrix $[\Sigma(\overrightarrow{k_{m}},E)]$ includes that due to the scattering of the electronic wavefunctions from the contacts into the device region, denoted by $\Sigma_L(\overrightarrow{k_{m}},E)+\Sigma_R(\overrightarrow{k_{m}},E)$ as well as the scattering of electronic wavefunctions inside the device due to incoherent processes such as phonons and non-idealities, denoted by  $\Sigma_s(\overrightarrow{k_{m}},E) $ (detailed in the supplementary information). The calculation of the in-scattering and the out-scattering functions involve a self consistent procedure, (detailed in the supplementary information), with the electron and the hole density operators $G^n(\overrightarrow{k_{m}},E)$, $G^p(\overrightarrow{k_{m}},E)$ defined as
\begin{eqnarray}
G^n(\overrightarrow{k_{m}},E)=G(\overrightarrow{k_{m}},E)\Sigma^{in}(\overrightarrow{k_{m}},E)G^{\dagger}(\overrightarrow{k_{m}},E), \nonumber \\
G^p(\overrightarrow{k_{m}},E)=G(\overrightarrow{k_{m}},E)\Sigma^{out}(\overrightarrow{k_{m}},E)G^{\dagger}(\overrightarrow{k_{m}},E). 
\end{eqnarray}
Upon convergence of the self-consistent quantities, the charge and heat currents are evaluated in the lattice basis as:
\begin{widetext}
\begin{align}
I^{j\rightarrow j+1}_C & =\underset{k_m}{\sum}i\frac{e}{\pi \hslash} \int[
G^n_{j+1,j}(\overrightarrow{k_{m}},E)H_{j,j+1}(E)   
-H_{j+1,j}(E)G^n_{j,j+1}(\overrightarrow{k_{m}},E) ]dE, \\
I_Q^{j\rightarrow j+1} & =\underset{k_m}{\sum}\frac{i}{\pi \hslash} \times  \int(E-\mu_H)[
G^n_{j+1,j}(\overrightarrow{k_{m}},E)  H_{j,j+1}(E)-H_{j+1,j}(E)
G^n_{j,j+1}(\overrightarrow{k_{m}},E) ]dE,  
\label{eq:heatcurrentnegf}
\end{align}
\end{widetext}
where $\mu_H$ is the electrochemical potential of the hot contact, $M_{i,j}$ is a generic matrix element of the concerned operator between two lattice points $i$ and $j$. In the tight-binding scheme used here, we only consider the next nearest neighbor such that $j=i \pm 1$. A list of parameters used for the simulation of the NEGF equations are given in Tab. I. \\
\begin{table}[]
\caption{Parameters used for simulation in this paper.}
\begin{center}
    \begin{tabular}{| p{3cm} |p{3cm}|}
    \hline
    \textbf{Parameters} & \textbf{Values}   \\ \hline \hline
    $T_H~~(k_BT_H)$  & $330K~~(28.435~meV)$  \\ \hline
      Length of device  & $20nm$ \\ \hline
    $T_C~~(k_BT_C)$ & $300K~~(25.85~meV)$  \\ \hline
     $D_O$ (SI) & $0.2F~eV^2$  \\ \hline
    $T=\frac{T_H+T_C}{2}~~(k_BT)$  & $315K~~(27.1425~meV)$   \\ \hline
      $S(E)$ (SI) & $\frac{2\pi D_O}{\hslash} \left(10\frac{E}{e}\right)^{-u}~eV^2$ \\ \hline
    $m_l$ & $m_e$    \\ \hline
     $a$ (lattice constant)& $2.7 \AA$ \\ \hline
    $m_t$ & $0.2m_e$   \\ \hline
     $E_c$  & $0eV$ \\ \hline

    \end{tabular}
       \vspace{1ex}

\end{center}
\justifying{\scriptsize \emph{\textbf{Note:}} ``SI"=supplementary information. $F=\frac{1}{N_xN_y}$, $N_x$ and $N_y$ being the number of lattice points in the $x$ and $y$ directions with electronic transport being in the $z-$direction. The parameter $u$ is related to the order of scattering process `$-r$' by the equation $u=-r+\frac{1}{2}$ for nanowires and $u=-r-\frac{1}{2}$ for bulk generators. $u=0$ for acoustic phonon scattering. $m_e$ is the free electron mass and $D_O$ is related to the acoustic deformation potential (see supplementary information). }
\end{table}

\section*{Results}

\begin{figure*}[!htb]

\includegraphics[scale=1.3]{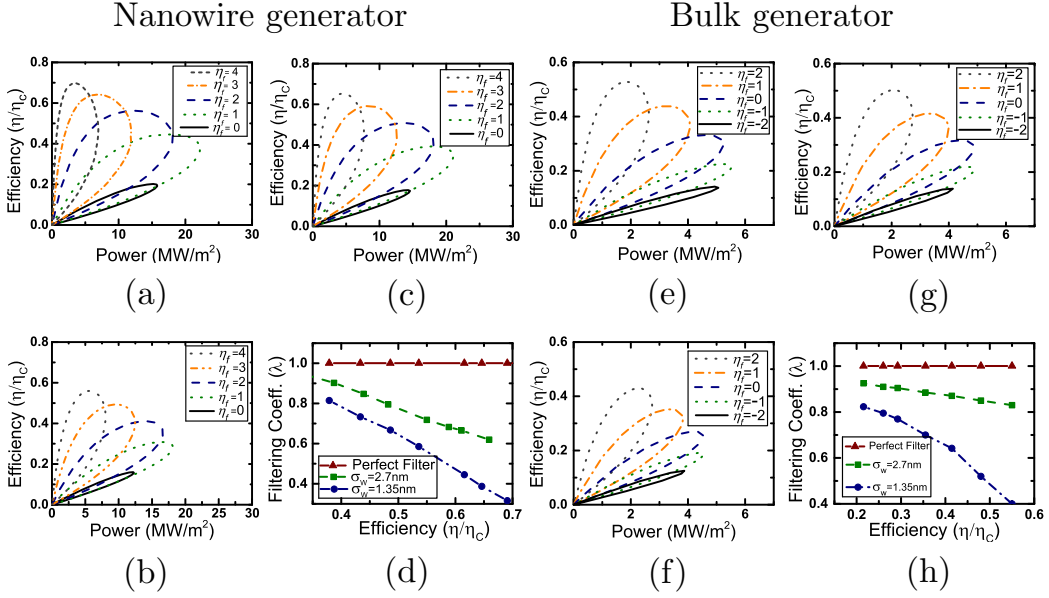}

\caption{Analysis of energy filtering  in the ballistic case. (a-d) Analysis for a $20nm$ long and  $2.7nm \times 2.7nm$ ballistic square  nanowire: (a) Power-efficiency trade-off analysis in case of Approach A for various values of $\eta_f=\frac{E_c-\mu}{k_BT}$. (b-c) Power-efficiency trade-off analysis in case of Approach B for various values of $\eta_f=\frac{E_c+E_b-\mu}{k_BT}$    (b) with a thin energy  barrier ($\sigma_w=1.35nm$, $E_b=150meV$), (c) with thick barrier ($\sigma_w=2.7nm$, $E_b=150meV$)  (d) Plot of filtering coefficient ($\lambda$) versus efficiency ($\eta/\eta_C$) for a thin, a thick and a perfect energy filtering barrier. (e-h) Analysis   for a $20nm$ long bulk generator: (e)  Power-efficiency trade-off analysis in case of   Approach A for various values of $\eta_f=\frac{E_c-\mu}{k_BT}$. (f-g) Power-efficiency trade-off analysis in case of  Approach B  for various values of  $\eta_f=\frac{E_c+E_b-\mu}{k_BT}$   (f) with thin barrier ($\sigma_w=1.35nm, ~E_b=150meV$), (g) with  a thick barrier ($\sigma_w=2.7nm$, $E_b=150meV$) (h) Plot of filtering coefficient ($\lambda$) versus efficiency ($\eta/\eta_C$) for the bulk generator in case of a thin, a thick and a perfect energy filtering barrier.}
\label{fig:ballistic_nanowire}
\end{figure*}

We now perform a detailed analysis of power generation for Approach A and Approach B using nanowire and bulk thermoelectric generators. The power-efficiency curves are plotted in the $\eta-P$ plane for several value of the reduced Fermi energy $\eta_f$ defined as:
\begin{equation}
\eta_f=
\left\{
	\begin{array}{ll}
		\frac{E_c-\mu}{k_BT}  & \mbox{for Approach A }  \\ \\
		\frac{E_c+E_b-\mu}{k_BT} & \mbox{for Approach B, } 
	\end{array}
\right.
\label{eq:eta_f}
\end{equation}
$E_b$ being the height of the  energy filtering barrier in case of Approach B. In all the discussions to follow, the efficiency of operation will be evaluated relative to the Carnot efficiency given by $\eta_C=1-T_C/T_H$.

\subsection*{Energy filtering in the ballistic limit}\label{coherent}
We start by considering nanowire and bulk thermoelectric generators in the ballistic or coherent limit. We plot, in Fig. \ref{fig:ballistic_nanowire}, the power density versus efficiency curves for a range of  $\eta_f$ for both Approach A and Approach B. In particular, Fig \ref{fig:ballistic_nanowire} (a)-(d) depict power generation characteristics for single-moded nanowire generators while Fig \ref{fig:ballistic_nanowire} (e)-(h) depict the same for bulk generators.   For both nanowire and bulk generators, the maximum power density  as well as the power density at a given efficiency for  Approach B increase with the  width of the energy barrier.  The peak performance in this case is achieved for a perfect energy filter with sharp transmission cut-off. This peak performance in Approach B is identical with that of Approach A. \\ 
\indent  The corresponding filtering coefficients for nanowire and bulk systems, plotted in Fig \ref{fig:ballistic_nanowire} (d) and (h) respectively, decrease in the high efficiency regime  which also  corresponds  to a  high value of $\eta_f$. This trend can be attributed to the smooth transmission cut-off at energies close to the barrier height $E_b$ (details in Supplemetary material). We hence conclude that in the ballistic limit, for both nanowire or bulk thermoelectric generators, the maximum power generation at a given efficiency is achieved via Approach A. This makes energy filtering via embedded nano barriers  of very limited use in the ballistic limit.   Hence, in the ballistic limit, when considering generators shorter than the mean free path, energy filtering with a single barrier (Approach B) does not provide any additional benefit over the traditional good thermoelectric generator (Approach A). \color{black} \\

\subsection*{Energy filtering in the diffusive  limit.} \label{incoherent}
\begin{figure*}[!hbt]
\hspace{-1cm} \includegraphics[scale=1.4]{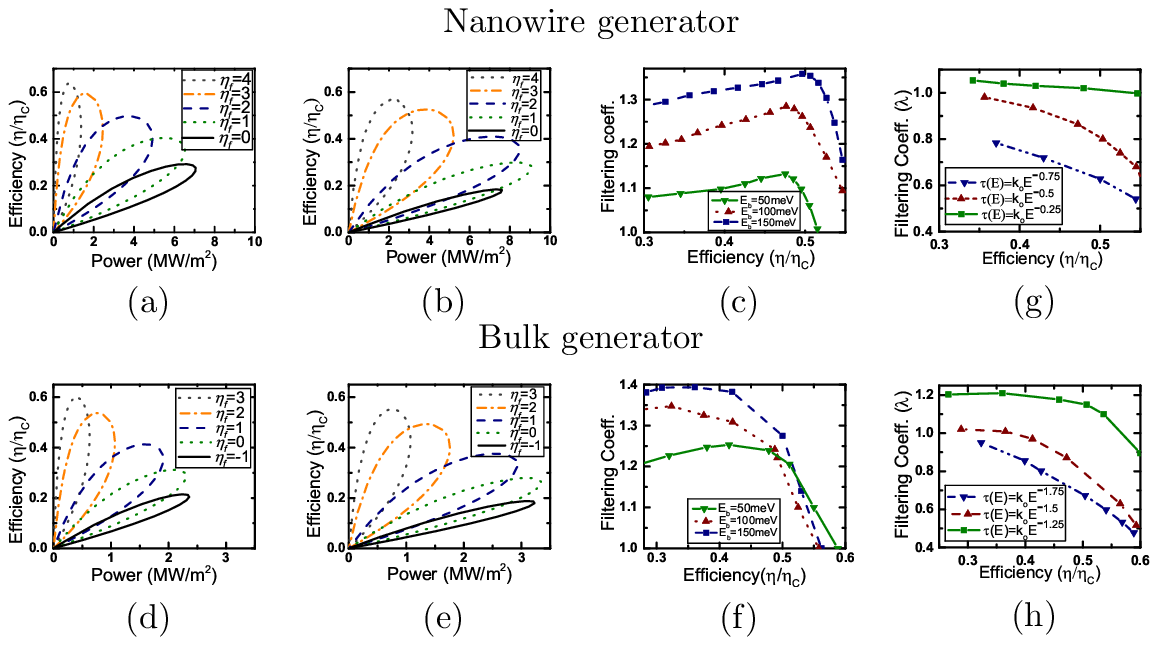}
\caption{Thermoelectric generation and filtering coefficient  analysis with incoherent scattering mechanisms (a-c) Analysis   for a square nanowire  of length $20nm$ and width $2.7nm$ with the inclusion of acoustic phonon scattering: (a) Power-efficiency trade-off analysis in case of Approach A for  various values of $\eta_f=\frac{E_c-\mu}{k_BT}$. (b) Power-efficiency trade-off analysis in case of  Approach B for various values of $\eta_f=\frac{E-E_c-E_b}{k_BT}$ ($E_b=150meV$ and $\sigma_w=2.7nm$). (c) Filtering coefficient plotted for various heights of the energy filtering barrier.  (d-f) Analysis   for a bulk generator of length $20nm$ with the inclusion of acoustic phonon scattering: (d) Power-efficiency trade-off analysis in case of Approach A for  various values of $\eta_f=\frac{E_c-\mu}{k_BT}$. (e)  Power-efficiency trade-off analysis in case of Approach B for various values of $\eta_f=\frac{E-E_c-E_b}{k_BT}$ ($E_b=150meV$ and $\sigma_w=2.7nm$). (f) Filtering coefficient plotted for various heights of the energy filtering barrier. (g-h) Filtering coefficients plotted for higher order scattering mechanisms with Gaussian energy barriers  ($\sigma_w=2.7nm$, $E_b=150meV$) for (g) nanowire generator with dimensions used in (a-c) and (h) bulk generator with dimensions used in (d-f).   }
\label{fig:nonballistic_nanowire}
\end{figure*}
\textbf{Energy filtering with acoustic phonon scattering:} We turn our attention to the diffusive limit with the inclusion of acoustic phonon scattering and plot in Fig. \ref{fig:nonballistic_nanowire}, the filtering analysis of the devices considered previously. In particular, Fig. \ref{fig:nonballistic_nanowire}~(a)-(c) demonstrates the generation characteristics for single-moded nanowires and Fig. \ref{fig:nonballistic_nanowire}~(d)-(f) demonstrates the power generation characteristics for bulk generators.   Contrary to the ballistic limit, acoustic phonon scattering  ensures an improved thermoelectric performance in Approach B rather than Approach A.  To explain the unusual trends noted in Fig. \ref{fig:nonballistic_nanowire}, specifically, the increase in filtering coefficient $\lambda$ for single-moded nanowire generators with increase in energy barrier height $E_b$ (Fig \ref{fig:nonballistic_nanowire} (c)),  we use the parameter $\Upsilon(E)$, which is identical to the Boltzmann transport coefficient (details given in supplementary information): \color{black}
 \[
 \Upsilon(E)=v_z^2(E)\tau(E)D_{1D}(E), 
 \]
 where $v_z(E)$, $\tau(E)$ and $D_{1D}(E)$ represent the electronic transport velocity, relaxation time and the 1-D density of states respectively. In case of perfect filtering or sharp transmission cut-off,  Approach B theoretically ensures an enhanced performance compared to Approach A, with the relative enhancement being an increasing function of the energy barrier height $E_b$, provided the parameter $\Upsilon$ is an increasing function of energy.  
 For bulk thermoelectric generators, under the assumption of uncoupled sub-bands and sharp transmission cut-off, we can define a parameter $\zeta(E)=\frac{ (E^{\frac{3}{2}}-E_b^{\frac{3}{2}} )E^r}{ \{(E-E_b)^{r+\frac{3}{2}} \}}$ with $E>E_b$ to access the relative enhancement in generated power via Approach B compared to Approach A (details given in supplementary information).  It can be shown that for acoustic phonon scattering, the parameter $\zeta>1$, which implies an enhanced generated power via Approach B (details given in supplementary information). \\
\indent   We thus note the non-trivial result that indicates an improvement of thermoelectric performance via energy filtering (Approach B) in the presence of acoustic phonon scattering.   Energy filtering is thus beneficial for devices dominated by ``acoustic phonon scattering"-like mechanisms ($\tau(E)\propto E^{\frac{1}{2}} $ for nanowires and $\tau(E)\propto E^{-\frac{1}{2}} $ for bulk) when they are longer than the mean free path. \color{black} We now explore what happens for the higher order scattering mechanisms. \\
  \textbf{Energy filtering with higher order scattering mechanisms:}
  A scattering mechanism, with relaxation time given by $\tau(E)=k_0E^r$, is said to be of order `$-r$',  with a lower value of the exponent $r$ denoting a higher order scattering process. For acoustic phonon scattering, the relaxation time is given by $\tau(E)\approx k_oE^{\frac{1}{2}}$ in the case of nanowires and $\tau(E) \approx k_oE^{-\frac{1}{2}}$ in the bulk case. 
Then, the question  arises, what is the effect of $r$ on the filtering coefficient?  Another related question is, if there is a minimum value of $r$, termed $r_{min}$,  beyond which energy filtering viz. Approach B ensures an enhanced performance compared to approach A?\\
\indent  As already discussed,   for single-moded nanowires, Approach B always ensures an enhanced thermoelectric performance when the parameter $\Upsilon(E)$ is an increasing function of energy (details given in supplementary information), provided that energy filtering is perfect.  Hence, in the case of non-parabolic bands or in the case where different scattering mechanisms dominate, the height of the energy barrier, for optimum performance, should be  approximately identical to the energy at which the rate of increase of $D\tau v^2_z$ saturates. The optimum doping for such a case should fix  the Fermi level close to the energy barrier height depending on the desired efficiency. \color{black} The rate of increase of $\Upsilon(E)$ with energy determines the relative advantage gained via Approach B, that is, for the same energy barrier height with $\tau(E)=kE^r$ and parabolic dispersion, filtering coefficient at a given efficiency is an increasing function of $r$.
 For a single-moded nanowire, $v_z^2(E)=\frac{2E}{m_l}$ and $D(E)=\frac{1}{\hslash\pi}\sqrt{\frac{m_l}{2E}}$, implying  $r_{min}=-\frac{1}{2}$. For $r<r_{min}$, Approach A leads to an optimized power generation degrading the filtering coefficient less than unity.  An analytical calculation of $r_{min}$ for bulk generators is not  trivial due to inter sub-band coupling.  However, assuming uncoupled sub-bands and perfect energy filtering, for  $E_b=150meV$, it can be shown that Approach B is advantageous for $r \gtrsim -0.7$ . (see supplementary information).  We hence note that in case of  devices longer than the mean free path, the benefit obtained from energy filtering decreases as the order of the dominating scattering mechanism  increases. \color{black}\\
\indent We plot, in Fig. \ref{fig:nonballistic_nanowire} (g) and (h), the filtering coefficient ($\lambda$) versus efficiency ($\eta/\eta_C$)  for nanowire and bulk generators affected by scattering mechanisms of order higher than that of acoustic phonon scattering.  The filtering coefficient vs. efficiency plots in  Fig. \ref{fig:nonballistic_nanowire} (h) indicate that for bulk thermoelectric generators, the calculated upper bound $r_{min}=-0.7$ under the simplified assumption of uncoupled sub-bands is indeed an overestimate.   To explain this behavior, we thus need to delve into the details of inter sub-band scattering and understand its contribution to power generation.   \\
\textbf{Role of intermode coupling in bulk generators:} We now uncover the role of coupling between the various transverse sub-bands  in the bulk case, in particular, the role of incoherent scattering in enhancing the filtering coefficient. Conservation of lateral momentum in the ballistic limit implies that, electrons from the high energy transverse sub-bands are reflected from the energy barrier  \cite{dattabook,LNE,Datta_Green,Vashaee_conserve,kim_conserve} and hence cannot contribute to the generated power. However, in  the diffusive limit, transverse sub-bands are coupled which may result in an electronic flow between them. The net scattering current between the sub-bands is however dictated by the relative non-equilibrium conditions of the respective sub-bands and  will  henceforth be referred to as the \emph{intermode coupling current}. Such a flow of the intermode coupling current between the sub-bands occurs in the region between the hot  contact and the barrier interface. This current subsequently flows to the cold contact and  can significantly enhance the generated power in the  case of energy filtering.\\
\begin{figure*}
 \centering
\includegraphics[scale=1.4]{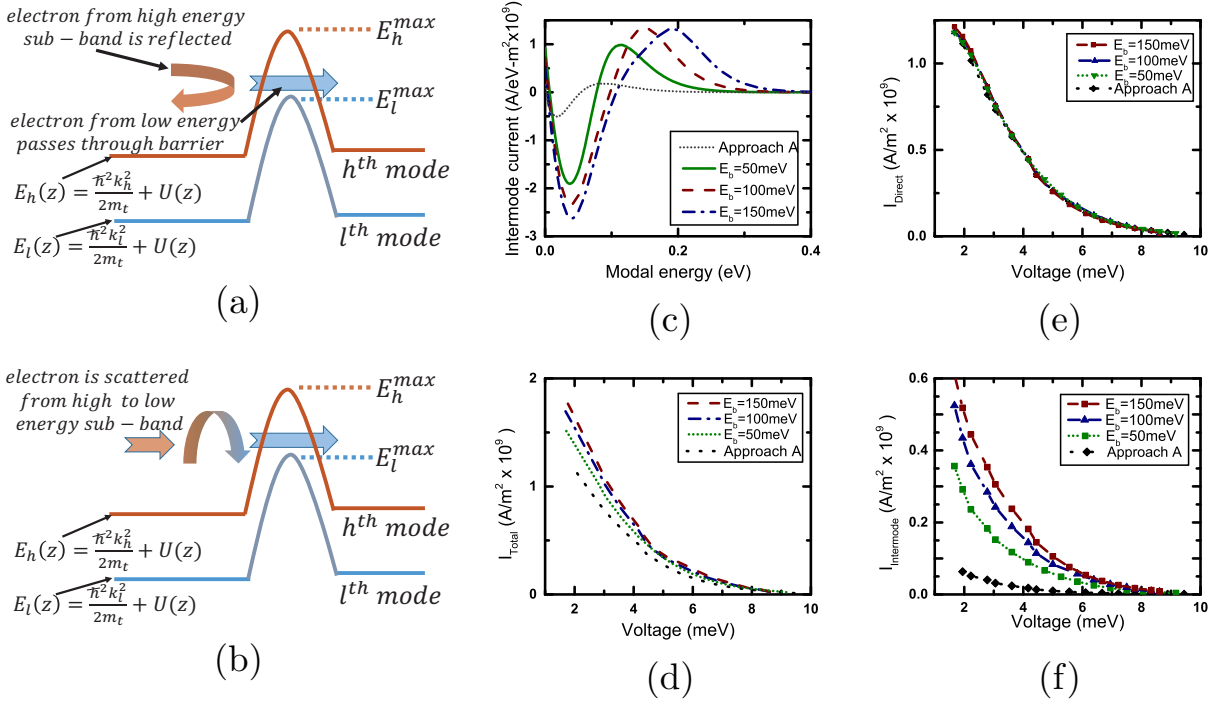}
\caption{Demonstration of performance improvement due  to inter sub-band coupling in energy filtering based bulk thermoelectric generators. (a,b) Schematic diagram illustrating electronic transport through a device with two sub-bands: the $l^{th}$ and the $h^{th}$ sub-band with lateral momentum $\hslash k_l$ and $\hslash k_h$ respectively. (a) Electronic transport when the sub-bands are uncoupled. In this case, the electrons from the $h^{th}$ sub-band are completely reflected from the barrier if the total energy of the electron is lower than $E_h^{max}$. (b) Electronic transport when the sub-bands are coupled. In this case, the electrons  from the $h^{th}$ sub-band (with energy between    $E_l^{max}$ and  $E_h^{max}$)   may be scattered to the low energy sub-bands between the hot contact and the barrier interface and subsequently flow towards the cold contact. (c) Plot of intermode coupling current vs. modal (sub-band) energy per unit area per unit energy  in case of  bulk generators at the maximum power.  (d-f) Current  profiles  per unit area at the maximum power at different bias voltages ($V$)  considering acoustic phonon scattering. Plot  of (d) the total current,    (e) the  current propagating directly between the contacts without changing sub-bands  and (f) the  intermode coupling current.  Simulations are done for a $20nm$ long device. For Approach B, an embedded Gaussian energy barrier is used ($\sigma_w=2.7nm$). An enhancement in generated power with barrier height results from an enhancement in the intermode coupling current.}
\label{fig:intermode2}
 \end{figure*}
\indent  The situation described above is schematically illustrated in Fig. \ref{fig:intermode2} (a) and (b) considering electronic transport through two sub-bands with transverse momentum $\hslash k_l$ and $\hslash k_h$. In the classical limit the approximate transmission cut-off of these modes are $E_l^{max}$ and $E_h^{max}$ respectively. 
Incoherent scattering can drive an electron (which would otherwise be reflected from the energy barrier)  with energy $E_l^{max}<E<E_h^{max}$ from the $h^{th}$ sub-band to the $l^{th}$ sub-band and contribute to the current flow.   For bulk thermoelectric generators the density of sub-bands $M(E)=\frac{m_t}{2\pi \hslash^2}(E-E_C)$  increases with  energy resulting in an increase in the intermode coupling current with the energy barrier height.\\
We plot in Fig. \ref{fig:intermode2} (c), the energy resolved  intermode coupling current  per unit area at the maximum power for various barrier heights taking acoustic phonon scattering into account. The electronic current flows out (positive value) of the higher energy sub-bands into (negative value) the lower energy modes.   We plot in Fig.  \ref{fig:intermode2} (d-f), the different current profiles at the maximum power at different voltage biases. The  total current per unit area ($I_{Total}$)  at the cold contact consists of two parts: (a) the direct current ($I_{Direct}$) that flows per unit area from the hot contact to the cold contact without changing sub-bands (b) the intermode coupling current $I_{Intermode}$ that flows per unit area from the higher energy sub-bands to the lower energy sub-bands between the hot contact and the energy barrier interface and consequently flows toward the cold contact.   Fig. \ref{fig:intermode2} (e) demonstrates  that $I_{Direct}$ remains almost unchanged with an increase in the height of the energy barrier. Hence, the  increase in the generated power with the height of the energy barrier in this case  is attributed to an increase in $I_{Intermode}$ as shown in Fig. \ref{fig:intermode2} (f).   Such an enhancement in generated power due to intermode coupling current is dependent on incoherent scattering and is absent in the  ballistic limit. \color{black}
\begin{figure*}[!htb]

\centering
\includegraphics[scale=1.3]{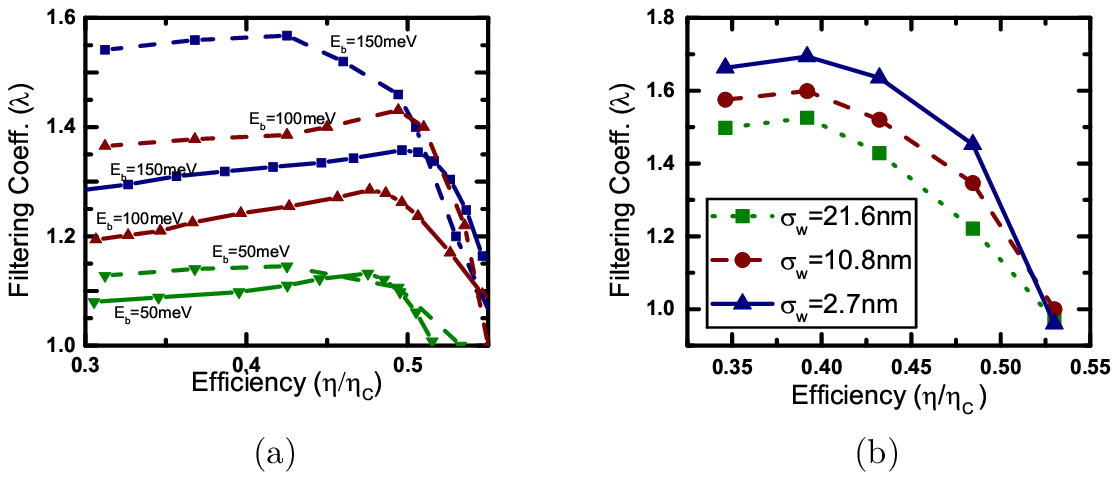}
\caption{Demonstration of the optimum conditions under which energy filtering should be employed. (a)  Demonstration of the enhancement in the filtering coefficient when the total length of the generator is much greater than the width of the energy filtering barrier. Plot of the filtering coefficient ($\lambda$) versus efficiency ($\eta/\eta_C$)  for  various barrier heights in case of $2.7nm$ wide square nanowires of length $20nm$ (solid curves) and $40nm$ (dashed curves). The nanowires are  embedded with a Gaussian energy barrier ($\sigma_w=2.7nm$). (b) Demonstration of deterioration of filtering coefficient when the width of the energy barrier is increased compared to the minimum width necessary to block an appreciable reverse flow of electrons from the cold contact. Filtering coefficient versus efficiency ($\eta/\eta_C$) for  a $170nm$ long diffusive nanowire with different energy  barrier widths (with $E_b=150meV$) taking acoustic phonon scattering into account. }
\label{fig:transmission}
\end{figure*}
\subsection*{Optimizing the filtering coefficient in the diffusive limit.}\label{optimize}
In this section, we reinforce that in the diffusive limit for $r>r_{min}$, the filtering coefficient  tends to maximize when the barrier width is `just sufficient' to block an appreciable reverse flow of electrons from the cold contact to the hot contact, a wider barrier than the optimized one being detrimental to the thermoelectric performance. \\
\textbf{Dependence of the filtering coefficient on device length:} For both nanowire and bulk generators with $r>r_{min}$, $\Upsilon(E)$ is minimum at the top of the energy barrier when energy filtering viz. Approach B is employed. 
Hence, Approach B is most useful when the device region between either of the contacts and the barrier interface is much longer compared to the width of the energy barrier. Such a design  facilitates most of the electronic transport in the region where the kinetic energy and $\Upsilon(E)$ are very high. This effect is demonstrated  in Fig.~\ref{fig:transmission} (a) where it is shown that for the same energy barrier width, the filtering coefficient increases for longer generators.\\ 
\textbf{Optimized energy barrier for maximizing the filtering coefficient:}
While it is obvious that a very thin energy barrier adversely affects the thermoelectric performance  due to reverse flow of electrons, a very wide barrier is also detrimental to the thermoelectric performance due to decrease in the parameter $\Upsilon(E)$ at the top of the barrier resulting in an overall decrease in the transmission probability (details in supplementary material).
This phenomenon is demonstrated in  Fig.~\ref{fig:transmission}~(b)   taking  incoherent (acoustic phonon) scattering into account  where it is shown that barriers wider than $2.7nm$ are detrimental to the thermoelectric performance.\\ 
\indent  Hence,  we conclude that for Approach B with $r>r_{min}$, the optimum energy filtering barrier is `just sufficiently wide' to block an appreciable reverse flow of electrons from the cold contact to the hot contact, the suitable generator length for employing energy filtering being at least  a few times greater than the optimum barrier width. \color{black}
\begin{figure*}[]
\centering
\includegraphics[scale=1.3]{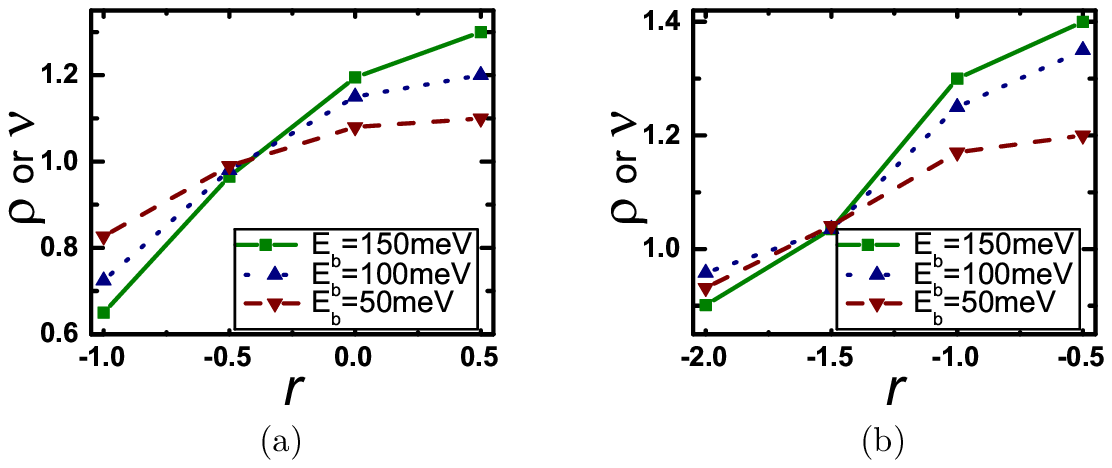}
\caption{Analysis of  performance improvement in thermoelectric generators via energy filtering under the condition $I_{Qp}>>I_{Qe}$ (heat conduction dominated by phonon). Plot of $\rho$ vs. $r$  for (a) a single moded square nanowire of length $20nm$  and width $2.7nm$ and (b) a  bulk generator of length $20nm$. The generators are embedded with a Gaussian energy barrier ($\sigma_w=2.7nm$). Plots are shown for three different energy barrier heights ($E_b$).}
\label{fig:phonon_heat}
\end{figure*}
\subsection*{Impact of phonon heat conduction}\label{phonon}
So far, in the above discussion, we have neglected the effect of phonon heat conduction. In realistic thermoelectric generators, however,  the efficiency of operation is often limited by phonon or lattice heat conduction\cite{phonon1,phonon2,phonon3,phonon4,phonon5,phonon6} if
\begin{equation}
I_{Qp}>>I_{Qe},
\label{eq:phn_heat}
\end{equation}
where $I_{Qp}$ and $I_{Qe}$ are the heat currents due to phonon and electronic conduction respectively at the hot contact.  The efficiency of operation in this case is given as:
\begin{equation}
\eta=\frac{P}{I_{Qp}+I_{Qe}}\approx \frac{P}{I_{Qp}}.
\label{eq:phn_eff}
\end{equation} \\
The expressions \eqref{eq:phn_heat} and \eqref{eq:phn_eff} naturally question the validity of our investigation where $I_{Qp}$ is neglected. \\
\indent Hence, we extend our proposed theory to the case where heat conduction is dominated by lattice heat conductance. In the case  where $I_{Qp}>>I_{Qe}$,  the operating point with maximum power becomes identical to the operating point with maximum efficiency. In this case, we define two  parameters which may be used to gauge the relative advantage gained via Approach B compared to Approach A. 
\[
\rho=\frac{P_B^{MAX}}{P_A^{MAX}}
\]
 \[
 \nu=\frac{\eta_{P_{MAX}}^B}{\eta_{P_{MAX}}^A} \approx \frac{P_B^{MAX}}{P_A^{MAX}},
 \]
 where $P_{A~(B)}^{MAX}$ and $\eta_{P_{MAX}}^{A~(B)}$ are the maximum power generated via Approach A(B) and the efficiency at the maximum generated power in case of Approach~A(B) respectively.\\
\indent We plot, in Fig. \ref{fig:phonon_heat} (a) and (b), $\rho$ vs. $r$ for nanowire and bulk   thermoelectric generators  respectively in the limiting case of $I_{Qp}>>I_{Qe}$ for various heights of the energy filtering barrier.  $\rho$ or $\nu$  increase monotonically with an increasing value of $r$.  In case of perfect filtering in nanowires, Approach B ensures an enhanced performance compared to Approach A in terms of maximum power generation when $\Upsilon(E)$ is an increasing function of energy. For such cases, both $\rho$ and $\nu$ increase with the height of the energy barrier. This general trends in $\rho$ or $\nu$ for nanowires  are also manifested in bulk generators for the case $I_{Qp}>>I_{Qe}$.  We hence note that preceding discussion on energy filtering is valid even  in the limit where lattice heat conduction dominates. \color{black}

\section*{Discussion}\label{conc}
So far, we have discussed the general conditions under which energy filtering enhances generated power in nanoscale and bulk thermoelectric generators. To demonstrate such conditions we have assumed smooth Gaussian barriers.  However our discussion is valid for energy filtering achieved via other means, for example, embedding nanoinclusions randomly throughout the generator.  Although disproven theoretically \cite{theory2}, such nanoinclusions are generally thought  to create resonant states which can further aid in enhancing the efficiency of thermoelectric generation. \color{black} In realistic devices, a number of electronic scattering mechanisms may be dominant such that the relaxation time of the electrons is a polynomial function of kinetic energy ($E$). 
\[
\tau(E)\approx\sum_i k_iE^{r_i}
\]
 We split the contributions $r_i$ into two groups (a) $r_i\geq r_{min}$ (b) $r_i<r_{min}$. With energy filtering, thermoelectric power generation is enhanced in the presence of the scattering mechanisms satisfying $r_i\geq r_{min}$  while the same is degraded in the presence of the scattering mechanisms satisfying $r_i<r_{min}$. In  a case where both types of scattering mechanisms are dominant or the energy band is non-parabolic, there is an optimum height of the energy filtering barrier at which the enhancement of generated power  is maximum in the case of Approach B\cite{theory1}.   The optimum barrier height in such a case should approximately be identical to the energy at which the parameter $D\tau v^2_z$ starts to saturate, the optimum position of the Fermi energy being in the range of a few $k_BT$ with respect to the top of the energy barrier depending on the desired efficiency. \color{black}  Our discussion in this paper, although general, should provide a qualitative  idea on the physics of the enhancement in generated power via energy filtering   in the ballistic and the diffusive limit and hence should provide general optimization guidelines for thermoelectric generators with material specific properties. 

\section*{Acknowledgements}
This work was partly supported by the IIT Bombay SEED grant and the Indian Space Research Organization RESPOND grant. The authors acknowledge Gang Chen for valuable suggestions and S. D. Mahanti for insightful discussions.
\section*{Author contributions statement}
A.S. conceived the idea. A.S. and  B.M.  contributed to the mathematical analysis, simulations,
development of the blue prints and the writing of the paper. A.S. and B.M.  extensively discussed
the science, the simulations and the results obtained.
\section*{Competing financial interests} The authors declare no competing financial interests.

\appendix

\section{NEGF equations for electronic transport mediated by incoherent scattering along with intermode coupling.}\label{appendix1}
In case of non-dissipative transport in nano devices, the generalized equations  for Green's function  and scattering matrix are  given by:
\begin{gather} 
G(\overrightarrow{k_{m}},E_z)=[E_zI-H-U-\Sigma(\overrightarrow{k_{m}},E_z)]^{-1} \nonumber \\
\Sigma(\overrightarrow{k_{m}},E_z)=\Sigma_L(\overrightarrow{k_{m}},E_z)+\Sigma_R(\overrightarrow{k_{m}},E_z)+\Sigma_s(\overrightarrow{k_{m}},E_z) \nonumber \\
 A(\overrightarrow{k_{m}},E_z)=i[G(\overrightarrow{k_{m}},E_z)-G^{\dagger}(\overrightarrow{k_{m}},E_z)] \nonumber \\
 \Gamma(\overrightarrow{k_{m}},E_z)=[\Sigma(\overrightarrow{k_{m}},E_z)-\Sigma^{\dagger}(\overrightarrow{k_{m}},E_z)], \nonumber \\
 \label{eq:negf}  
\end{gather}

where $H$ is the discretized Hamiltonian matrix (constructed using  the effective mass approach \cite{dattabook}), $U=-eV$ is the electronic potential energy in the band and $\Sigma_L(\overrightarrow{k_{m}},E_z)+\Sigma_R(\overrightarrow{k_{m}},E_z)$ and $\Sigma_s(\overrightarrow{k_{m}},E_z)$ describe the effect of  coupling and scattering of electronic wavefunctions  due to contacts and electron-phonon interaction respectively.
The device Hamiltonian, constructed using the effective mass approach, is given by \cite{dattabook}:

\[
\left [ H \right ] =\begin{bmatrix}
-2t_0 & t_0  & 0 & 0 & \dots & \dots & \dots  \\
t_0 & -2t_0 & t_0 &  0  & \dots  & \dots & \dots \\
0 &t_0 & -2t_0 & t_0 & 0  & \dots & \dots \\
\vdots \ddots & \ddots  & \ddots & \ddots   & \ddots &  \ddots \\
\dots & \dots & 0 &  t_0 & -2t_0 & t_0 & 0  \\
\dots & \dots & \dots & 0 &  t_0 & -2t_0 & t_0  \\

\dots & \dots & \dots & 0 & 0 &  t_0 & -2t_0   \\
\end{bmatrix},
\]

 where $t_0$ is the nearest neighbour tightbinding parameter constructed using effective mass approximation \cite{dattabook}.
 \[
 t_0=\frac{\hslash^2}{2m_la^2},
 \] 
$a$ being the effective lattice constant used for simulation given in Tab.~I.  In the above sets of equations, $\overrightarrow{k_{m}}$ denote the transverse wave-vector and $E_z$ is the free variable denoting the energy of the electrons  along the transport direction. $A(\overrightarrow{k_{m}},E_z)$ is the $1-D$ spectral function  for the $m^{th}$ sub-band and $\Gamma(\overrightarrow{k_{m}},E_z)$ is the broadening matrix for the $m^{th}$ sub-band at longitudinal energy $E_z$. 
For moderate electron-phonon interaction, it is generally assumed that the real part of $\Sigma_s=0$. Hence, 
\begin{equation}
\Sigma_s(\overrightarrow{k_{m}},E_z)=i\frac{ \Gamma_s(\overrightarrow{k_{m}},E_z)}{2}=\Sigma^{in}_s(\overrightarrow{k_{m}},E_z)+\Sigma^{out}_s(\overrightarrow{k_{m}},E_z)
\label{eq:sigma_phonon}
\end{equation}
$ \Sigma^{in}(\overrightarrow{k_{m}},E_z)$ and $ \Sigma^{out}(\overrightarrow{k_{m}},E_z)$ are the inscattering and the  outscattering functions which models the rate of scattering of the  electrons from the contact and inside the device.

\begin{multline}
  \Sigma^{in}(\overrightarrow{k_{m}},E_z)=\Sigma^{in}_L(\overrightarrow{k_{m}},E_z)+\Sigma^{in}_R(\overrightarrow{k_{m}},E_z) \\
  +\Sigma^{in}_s(\overrightarrow{k_{m}},E_z) \nonumber 
  \end{multline}
  \begin{multline}
   \Sigma^{out}(\overrightarrow{k_{m}},E_z)=\Sigma^{out}_L(\overrightarrow{k_{m}},E_z)+\Sigma^{out}_R(\overrightarrow{k_{m}},E_z) \\
   +\Sigma^{out}_s(\overrightarrow{k_{m}},E_z), \nonumber 
\label{eq:sig}
\end{multline}
where the subscript $'L'$, $'R'$ and $'s'$ denote the influence of left contact, right contact and electron-phonon scattering respectively. The in-scattering and out-scattering functions are dependent on the contact quasi-Fermi distribution functions as:
   
\begin{gather}
 \Sigma^{in}(\overrightarrow{k_{m}},E_z)=\underbrace{\Gamma_L(E_z)f_L(E_z+\frac{\hslash^2 \overrightarrow{k_m}^2}{2m_t})}_{inflow~from~left~contact} \nonumber \\+\underbrace{\Gamma_R(E_z)f_R(E_z+\frac{\hslash^2 \overrightarrow{k_m}^2}{2m_t})}_{inflow~from~right~contact}+\underbrace{\Sigma^{in}_s(\overrightarrow{k_{m}},E_z)}_{inflow~due~to~phonons} \nonumber  \\
 \Sigma^{out}(\overrightarrow{k_{m}},E_z)=\underbrace{\Gamma_L(E_z)\Big\{ 1-f_L(E_z+\frac{\hslash^2 \overrightarrow{k_m}^2}{2m_t})\Big \} }_{outflow~to~left~contact} \nonumber \\+\underbrace{\Gamma_R(E_z)\Big \{ 1-f_R(E_z+\frac{\hslash^2 \overrightarrow{k_m}^2}{2m_t})\Big \}}_{outflow~to~right~contact}+\underbrace{\Sigma^{out}_s(\overrightarrow{k_{m}},E_z)}_{outflow~due~to~phonons}, \nonumber  \\
 \label{eq:sig1}
 \end{gather}
where $f_{L(R)}$ denote the quasi-Fermi distribution of left(right) contact.
The rate of scattering of electrons due to phonons is dependent on the electron and the hole correlation functions $(G^n$ and $G^p)$ and is given by:

\begin{gather}
\Sigma^{in}_s(\overrightarrow{k_{m}},E_z)=D_O
\underset{\overrightarrow{q_{t}}}{\sum}G^n(\overrightarrow{k_{m}}+\overrightarrow{q_{t}},E_z-\Delta E_{\overrightarrow{k_{m}}+\overrightarrow{q_{t}},\overrightarrow{k_{m}}}) \nonumber \\
\Sigma^{out}_s(\overrightarrow{k_{m}},E_z)=D_O
\underset{\overrightarrow{q_{t}}}{\sum}G^p(\overrightarrow{k_{m}}+\overrightarrow{q_{t}},E_z-\Delta E_{\overrightarrow{k_{m}}+\overrightarrow{q_{t}},\overrightarrow{k_{m}}}),
\label{eq:sigma}
\end{gather}

where $\Delta E_{\overrightarrow{k_{m}}+\overrightarrow{q_{t}},\overrightarrow{k_{m}}}=\frac{\hslash^2 (\overrightarrow{k_{m}}+\overrightarrow{q_{t}})^{2}}{2m_t}-\frac{\hslash^2 \overrightarrow{k_{m}}^{2}}{2m_t}$.

 $G^n(\overrightarrow{k_{m}},E_z)$ and $G^p(\overrightarrow{k_{m}},E_z)$ are the electron and the hole correlation functions for the $m^{th}$ sub-band and $\{\overrightarrow{q_t}\}$ denotes the set of transverse phonon wave vectors. The electron and the hole correlation functions are again related to the  electron in-scattering and the electron out-scattering functions via the equations:
\begin{gather}
G^n(\overrightarrow{k_{m}},E_z)=G(\overrightarrow{k_{m}},E_z)\Sigma^{in}(\overrightarrow{k_{m}},E_z)G^{\dagger}(\overrightarrow{k_{m}},E_z) \nonumber \\
G^p(\overrightarrow{k_{m}},E_z)=G(\overrightarrow{k_{m}},E_z)\Sigma^{out}(\overrightarrow{k_{m}},E_z)G^{\dagger}(\overrightarrow{k_{m}},E_z) \nonumber \\
 \label{eq:correlation} 
\end{gather}
 Solving the  dynamics of the entire system involves a self consistent solution of \eqref{eq:negf},  \eqref{eq:sigma_phonon}, \eqref{eq:sigma} and  \eqref{eq:correlation}.
 For momentum scattering due to acoustic phonons, $D_O$ in the above equations can be related to the acoustic phonon deformation potential ($D_{ac}$) by:

\begin{equation}
D_O=\frac{ D_{ac}^2k_BTF}{\rho v_s^2 a^3},
\end{equation}

where $F$ is known as the form factor and denotes the spacial spread of the phonon wave-vectors. $\rho$ and $v_s$ denote the mass density and the velocity of sound in the material respectively. For the purpose of our simulation, we have used the parameters of bulk silicon. 
  The spectral function for the $m^{th}$ sub-band is given by 
\[
A(\overrightarrow{k_{m}},E_z)=G^n(\overrightarrow{k_{m}},E_z)+G^p(\overrightarrow{k_{m}},E_z)
\]

The electron density and current at the grid point $j$ can be calculated from the above equations as:

\[
n_j=\underset{m}{\sum}\int\frac{[G^n(\overrightarrow{k_{m}},E_z)dE_z]}{\pi aA}
\]

\begin{eqnarray}
I^{j\rightarrow j+1}=\underset{k_m}{\sum}\frac{e}{\pi \hslash} Im\int[H_{j+1,j}(E_z)G^n_{j,j+1}(\overrightarrow{k_{m}},E_z) \nonumber \\
-G^n_{j+1,j}(\overrightarrow{k_{m}},E_z)H_{j,j+1}(E_z)]dE_z, \nonumber \\
\label{eq:currentnegf}
\end{eqnarray}

where $a$ is the distance between two adjacent grid points and $A$ is the cross sectional area of the device. $\hslash k_m$ denotes the transverse momentum of the electrons in the $m^{th}$ sub-band. The summations in \eqref{eq:currentnegf} run over all the sub-bands available for conduction.

The heat current flowing through the device is given by:

\begin{eqnarray}
I_Q^{j\rightarrow j+1}=\underset{k_m}{\sum}\frac{1}{\pi \hslash} \times (E_z+E_m-\mu_H) Im\int[H_{j+1,j}(E_z)\nonumber \\
G^n_{j,j+1}(\overrightarrow{k_{m}},E_z) 
-G^n_{j+1,j}(\overrightarrow{k_{m}},E_z)H_{j,j+1}(E_z)]dE_z, \nonumber \\
\label{eq:heatcurrentnegf}
\end{eqnarray}
where $E_z$ is the kinetic energy of the electrons due to momentum along the transport direction and  $E_m$ is the kinetic energy of the electron due to momentum in the transverse direction. 
 The intermode coupling current can be calculated from the phonon scattering matrices using the formula,
\begin{eqnarray}
I_{Intermode}=\frac{2e}{h} \sum_{\overrightarrow{k_{m}}}\int\Sigma^{out}_{s}(\overrightarrow{k_{m}},E_z)G^n(\overrightarrow{k_{m}},E_z) \nonumber \\
-\Sigma^{in}_{s}(\overrightarrow{k_{m}},E_z)G^p(\overrightarrow{k_{m}},E_z)dE_z ,\nonumber \\
\end{eqnarray}
\color{black}
\subsection*{Calculating the transmission probability from Green's function} 

\emph{\underline{For coherent transport}}: The case of coherent transport corresponds to $D_O=0$. In this case, the transmission coefficient can be calculated from the broadening matrices using the formula \cite{dattabook,LNE}

\begin{equation}
T(E_z)=Trace\left[ \Gamma_{L}(E_z)G(E_z) \Gamma_{R}(E_z)G^{\dagger}(E_z)\right].
\end{equation}
 
\emph{\underline{For incoherent transport:}} For incoherent transport, transmission coefficient is ill-defined. For devices dominated by elastic scattering mechanisms, we can however define the probability of transmission of an electron via the formula \cite{dattabook,LNE}

\begin{eqnarray}
T(E_z)=\frac{i}{f_L(E_z)-f_R(E_z)} [G^n_{j+1,j}(E_z)H_{j,j+1}(E_z)\nonumber \\
-H_{j+1,j}(E_z)G^n_{j,j+1}(E_z)],\nonumber \\
\label{eq:currentnegf}
\end{eqnarray}

where $j$ is any number between $1$ and $N_p-1$, $N_p$ being the number of lattice points along the transport direction.
\color{black}

\section{Some results for perfect filtering.}\label{appendix0}
\begin{figure}[h]
\subfigure[]{\hspace{-.5cm}\includegraphics[scale=.18]{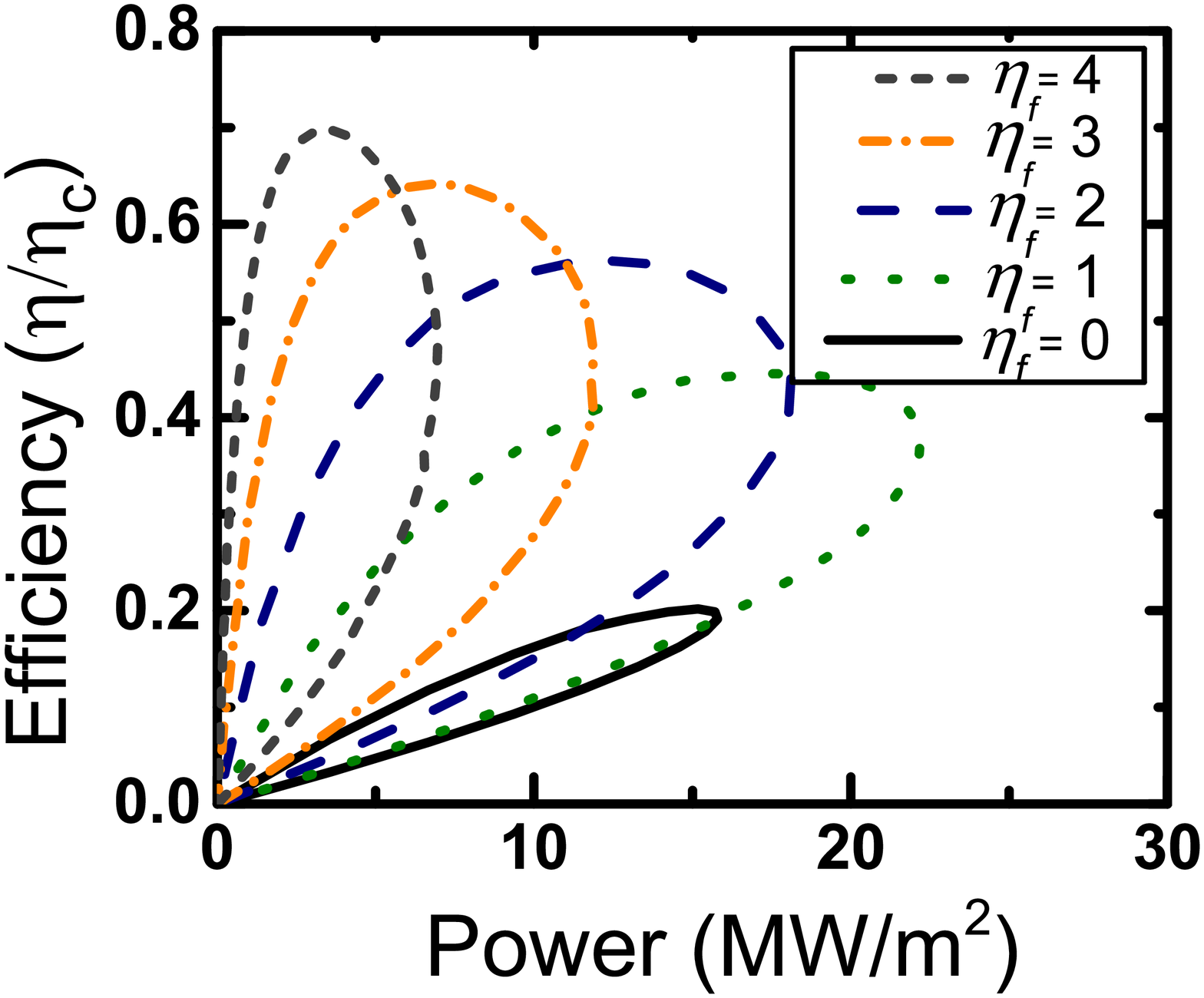}
}\subfigure[]{\hspace{-.6cm}\includegraphics[scale=.18]{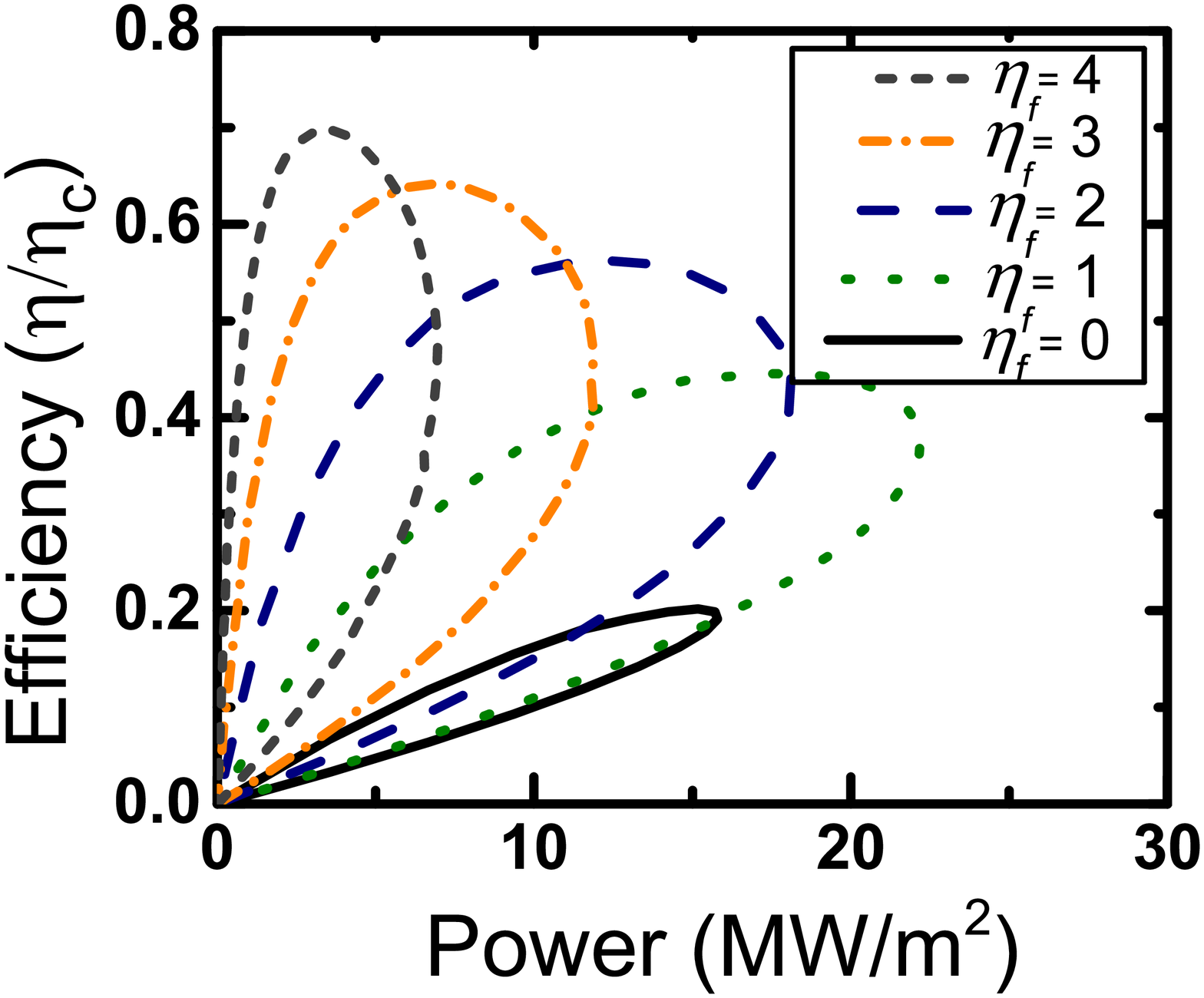}}
\subfigure[]{\hspace{-.5cm}\includegraphics[scale=.18]{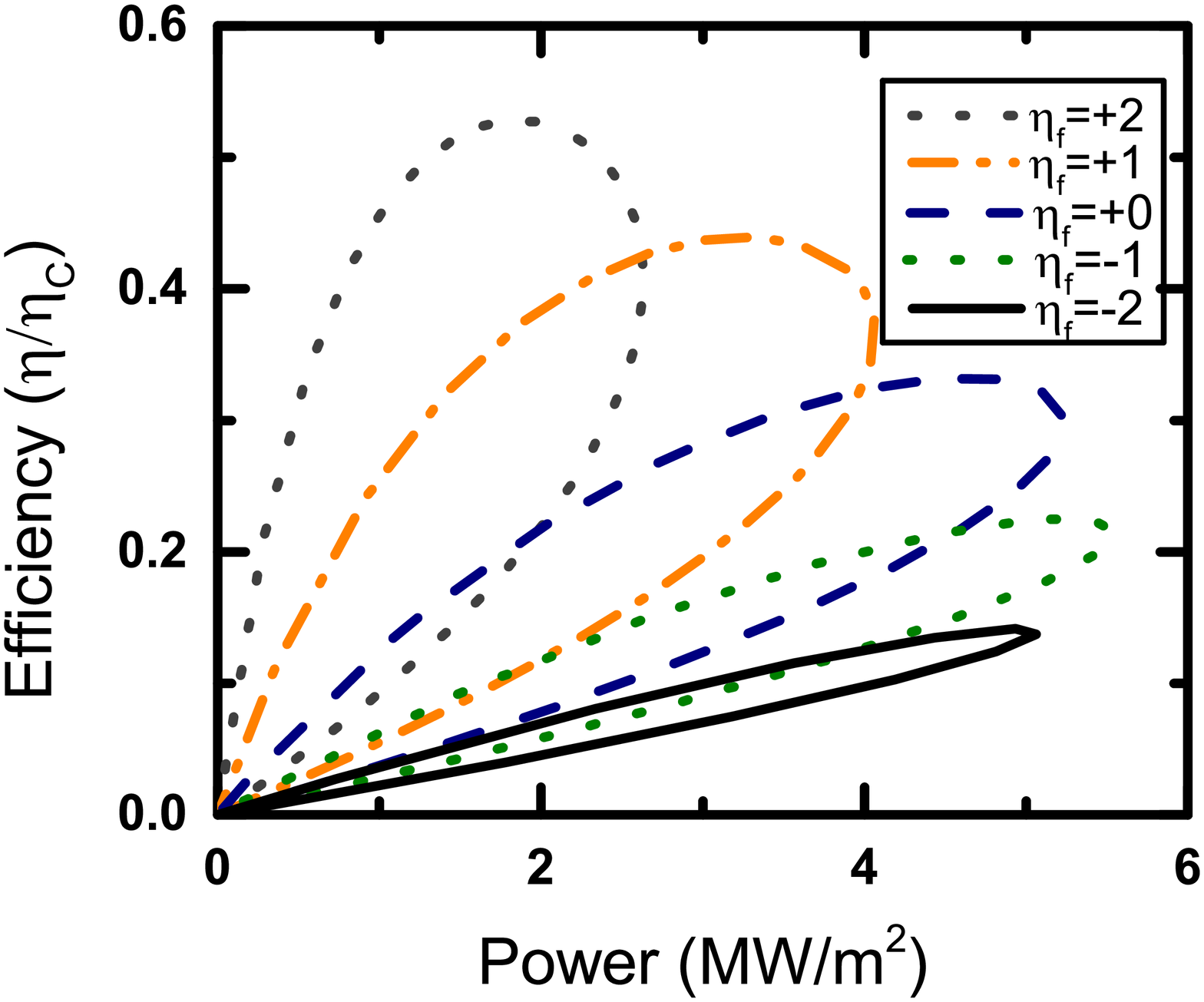}
}\subfigure[]{\hspace{-.5cm}\includegraphics[scale=.18]{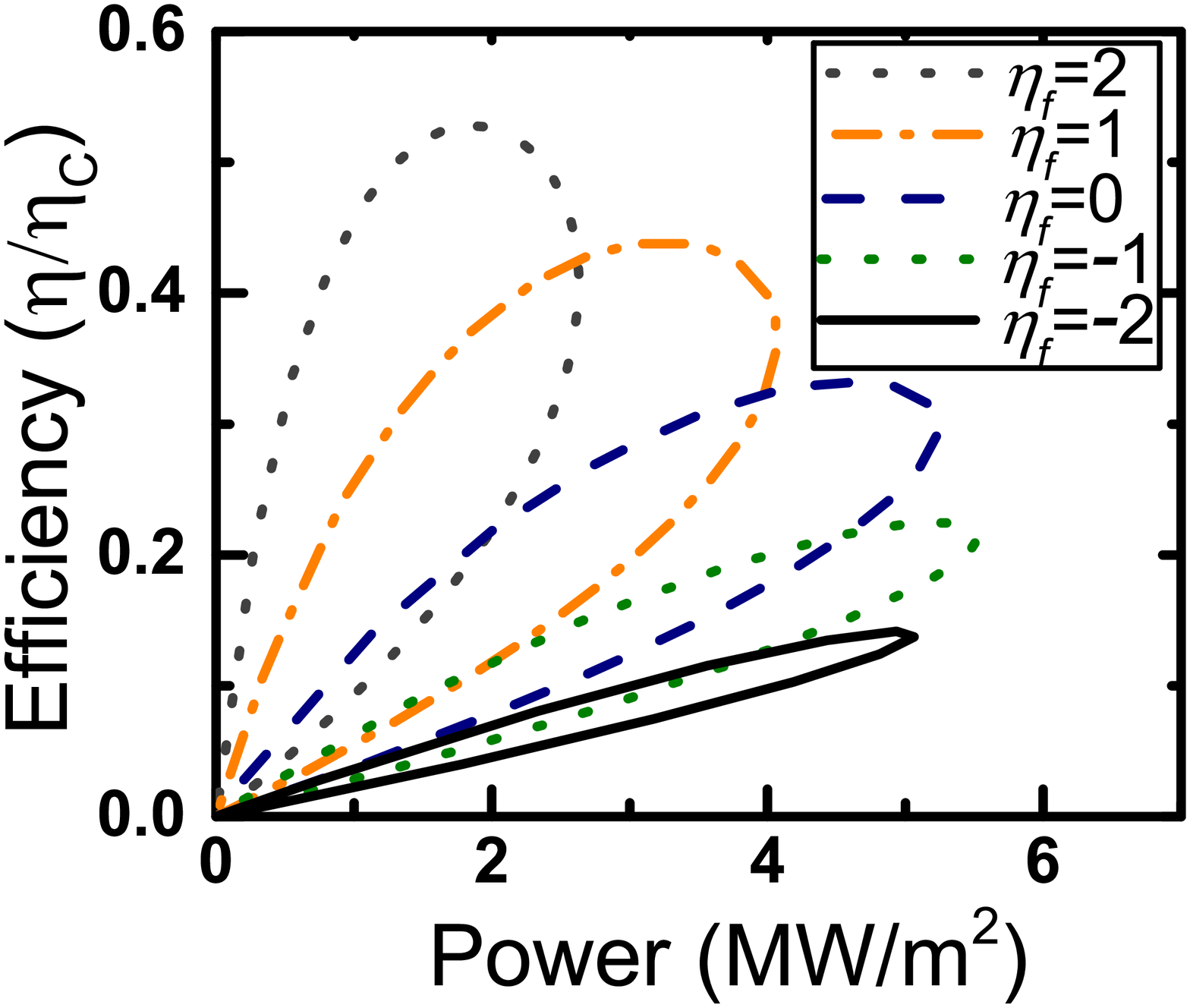}}
\subfigure[]{\hspace{-.6cm}\includegraphics[scale=.19]{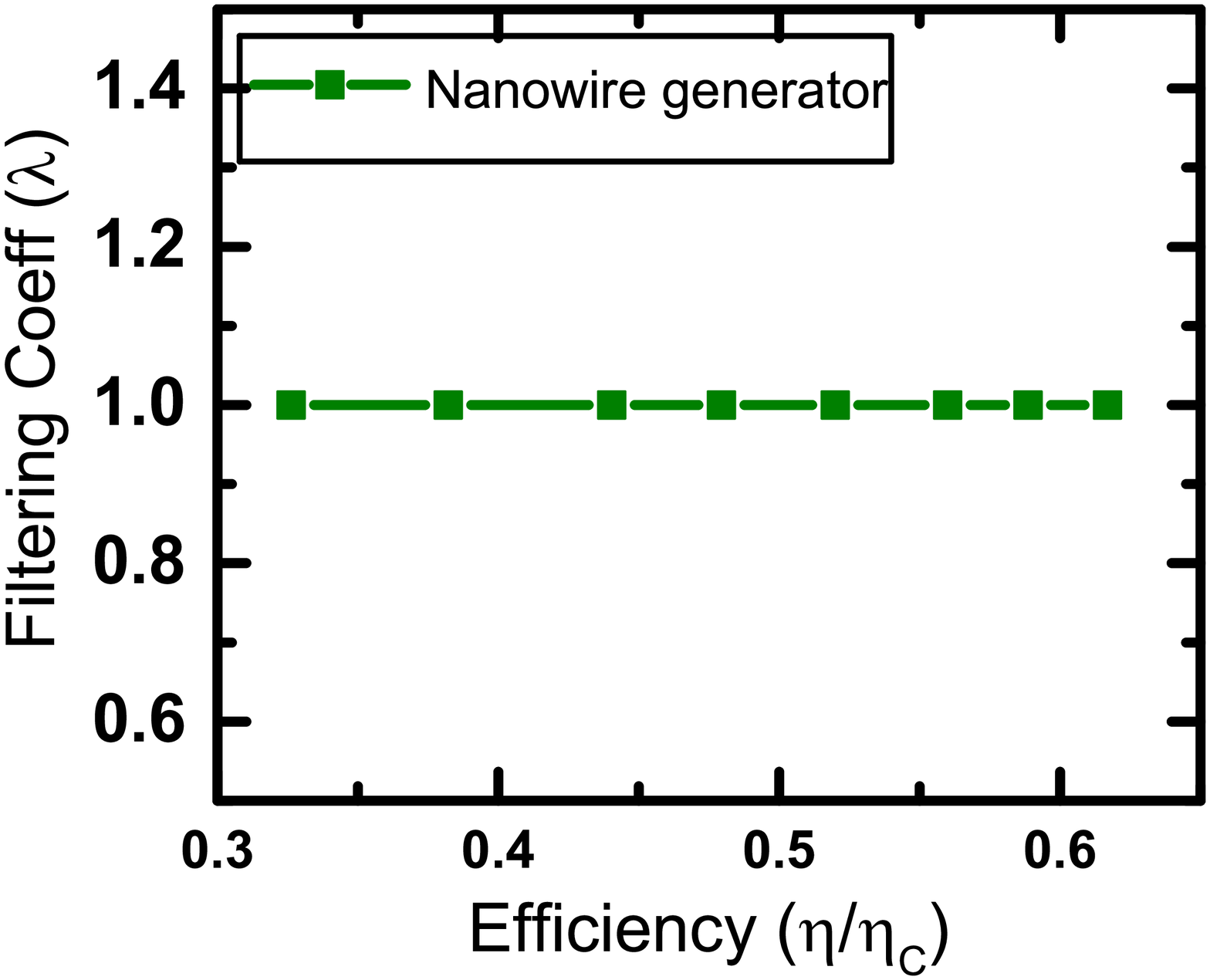}
}\subfigure[]{\hspace{-.5cm}\includegraphics[scale=.19]{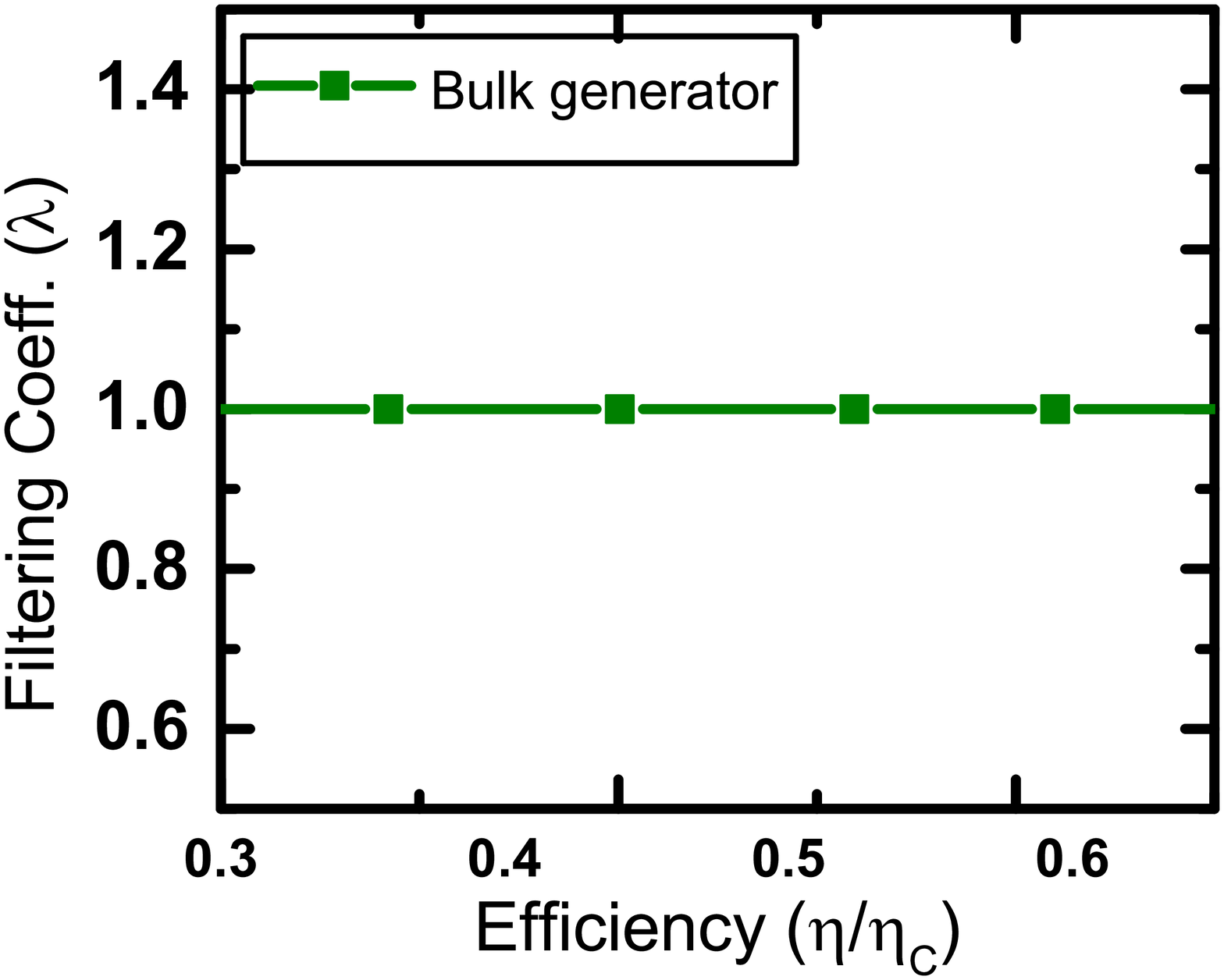}}
\caption{Power-efficiency trade-off and energy filtering analysis for ballistic devices. (a,b) Plot of power density versus efficiency  for a $20nm$ long and  $2.85nm \times 2.85nm$ ballistic square  nanowire generator (a) without  energy filtering, (b) with energy filtering due to conduction band-edge shifted to $E_c'=E_c+E_b$  (c, d)- Plot of power density versus efficiency  for  a ballistic bulk thermoelectric generator (d) without  energy filtering, (e) with energy filtering due to conduction band-edge shifted to $E_c'=E_c+E_b$. (e) Plot of  Filtering coefficient ($\lambda$) vs. efficiency $(\eta/\eta_C)$ for a ballistic nanowire in (b),  (f) Plot of  Filtering coefficient ($\lambda$) vs. efficiency $(\eta/\eta_C)$ for a ballistic bulk generator in (d). Simulations are done for $E_c=0$ and $E_b=  150meV$. $\eta_f=\frac{E_c+E_b-\mu}{kT}$}
\label{fig:ballistic_nanowire}
\end{figure} 
\begin{figure}[h]
\subfigure[]{\hspace{-.4cm}\includegraphics[scale=.185]{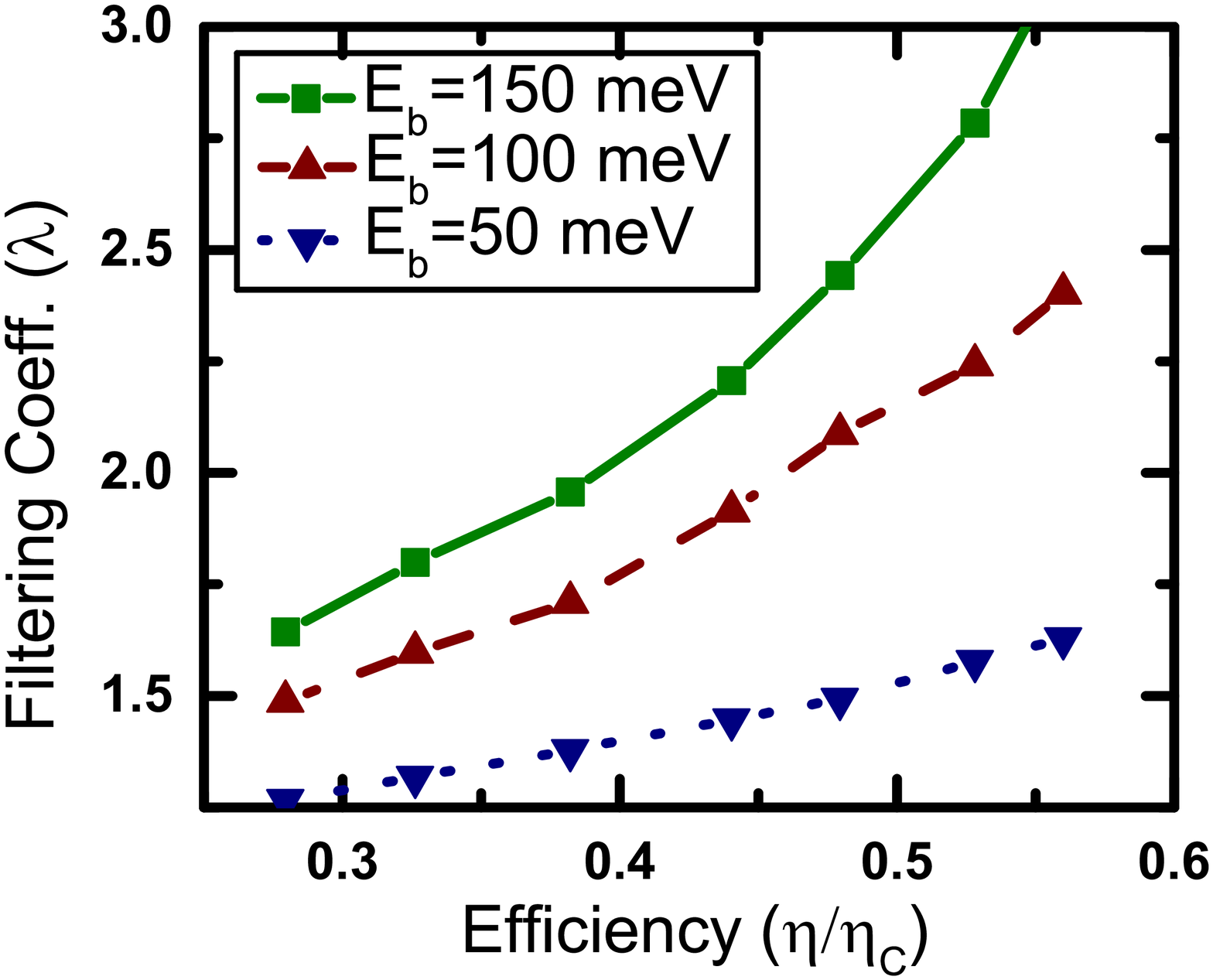}}%
\subfigure[]{\hspace{-.5cm}\includegraphics[scale=.185]{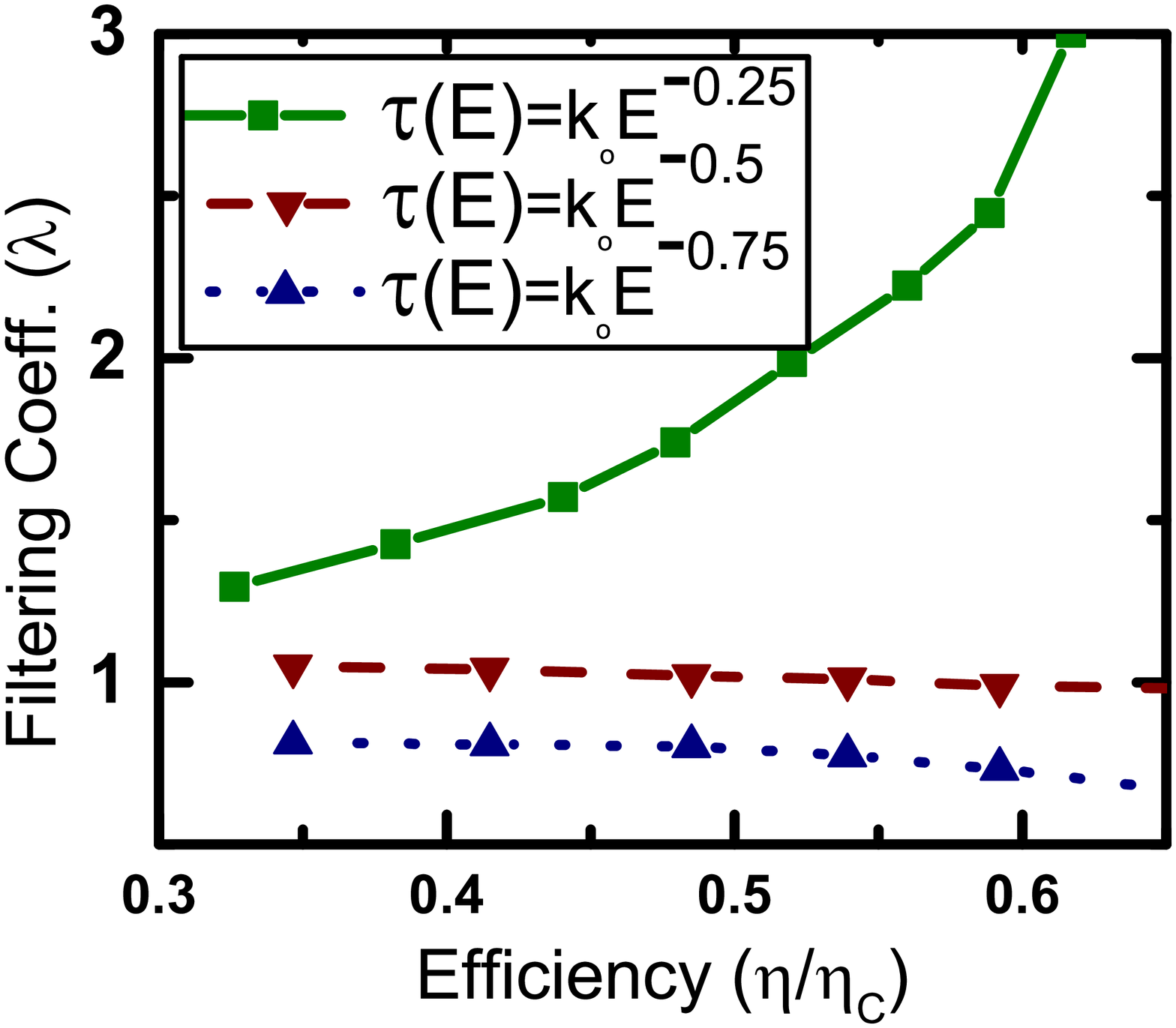}
}%
\caption{The case of perfect filtering for square nanowires of width $2.85nm$ and length $20nm$ . Plot of  (a) Filtering coefficient ($\lambda$) vs efficiency ($\eta/\eta_C$) for acoustic phonon scattering for three different cut-off energy and  (b)Filtering coefficient ($\lambda$) vs efficiency ($\eta/\eta_C$) for other incoherent scattering mechanisms with relaxation time $\tau$ given by $\tau(E)=k_oE^r$. Results are shown for $r=-0.25$, $r=-0.5$ and $r=-0.75$. A cut-off energy $E_b=150meV$ is used for the simulation. }
\label{fig:perfect_nanowires}
\end{figure}

\begin{figure}[h]
\subfigure[]{\hspace{-.4cm}\includegraphics[scale=.185]{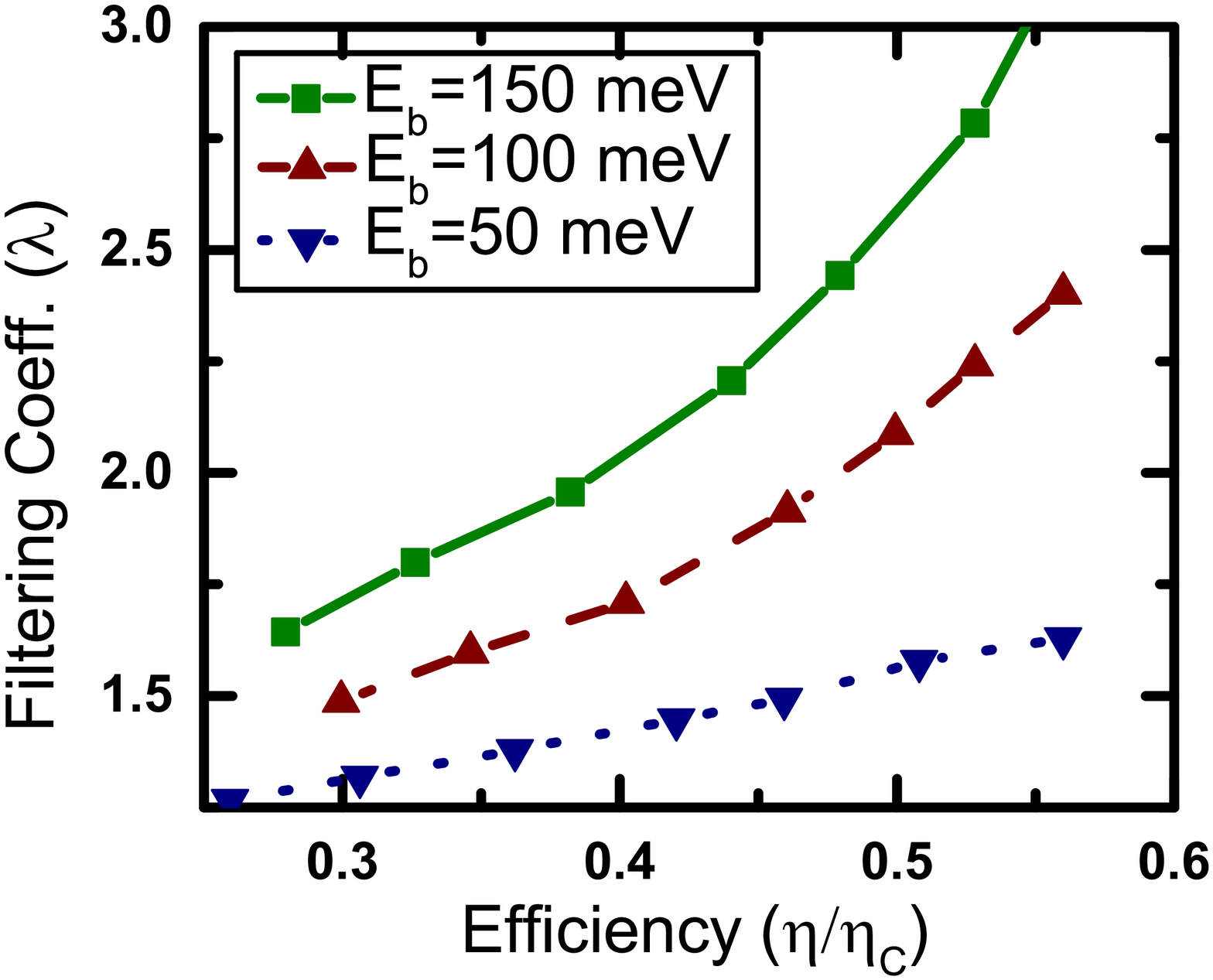}}%
\subfigure[]{\hspace{-.5cm}\includegraphics[scale=.185]{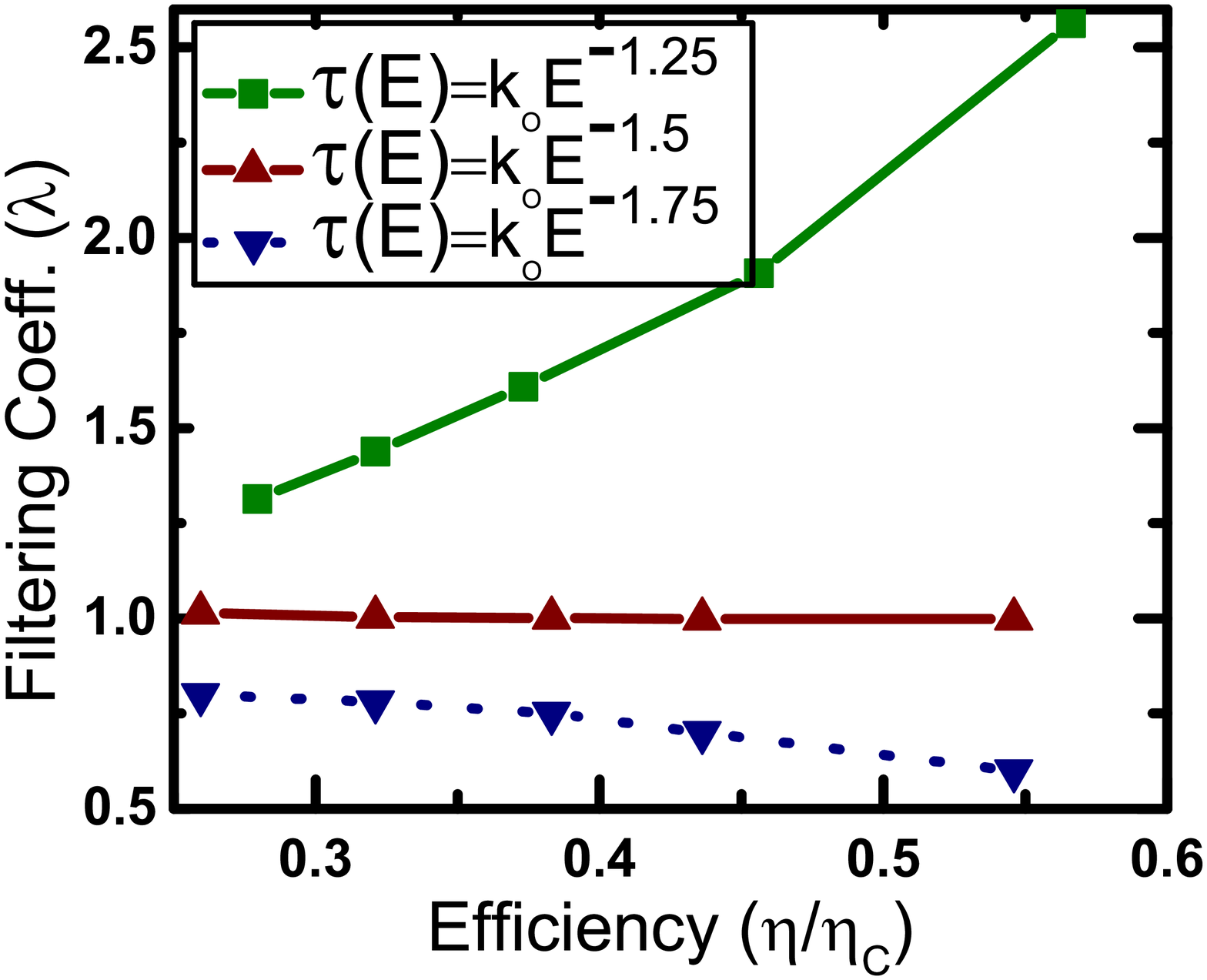}
}%
\caption{The case of perfect filtering for bulk generators with length $20nm$ . Plot of  (a) Filtering coefficient ($\lambda$) vs efficiency ($\eta/\eta_C$) for acoustic phonon scattering for three different cut-off energy and  (b)Filtering coefficient ($\lambda$) vs efficiency ($\eta/\eta_C$) for other incoherent scattering mechanisms with relaxation time $\tau$ given by $\tau(E)=k_oE^r$. Results are shown for $r=-1.25$, $r=-1.5$ and $r=-1.75$. A cut-off energy $E_b=150meV$ is used for the simulation. }
\label{fig:perfect_bulk}
\end{figure}
 In this section, we present some simulation results with perfect energy filtering for coherent and incoherent scattering.

{\emph{Perfect filtering for ballistic generators:}} Perfect filtering for ballistic transport is simulated by shifting the minimum potential of the conduction band  at a higher energy ($E_c'=E_c+E_b$) such that no electrons below $E_c'$ can contribute to the electrical current between the contacts. Fig. \ref{fig:ballistic_nanowire} compares the power-efficiency trade-off curves for ballistic nanowires and bulk thermoelectric generators with perfect filtering.

{\emph{Perfect filtering for generators dominated by incoherent scattering:}}
 To calculate  the results for perfect filtering in case of incoherent electronic transport, we simulate NEGF transport equations through  a device with conduction band edge $E_c=0$ while varying the Fermi potential and the voltage bias. The cut-off energy $E_b$ is chosen for such a device and only the electronic current above the cut-off energy is taken into consideration. In such a case, there is no effect of imperfect filtering by the energy barrier on the electronic transport.\\
 \indent Simulation results for perfect energy filterng for bulk and nanowire thermoelectric generators is presented in Fig. \ref{fig:perfect_nanowires} and \ref{fig:perfect_bulk} respectively. For thermoelectric generators dominated by incoherent scattering, the filtering coefficient ($\lambda$) increases monotonically with efficiency for $r>r_{min}$. This phenomenon mainly occurs due to the low conductivity at the edge of the conduction band for $r>r_{min}$. In the high efficiency regime, only the electrons at the conduction band edge contributes to the electrical current. As a result, the electrical current decreases rapidly in devices without energy filtering in the high efficiency regime of operation. The decrease in electrical current for generators with energy filtering is not so rapid resulting in an increase in the filtering coefficient in the high efficiency regime of operation. 

\color{black}

\section{Derivation of scattering self-energies for higher order local scattering mechanisms.} \label{appendix2}

 For elastic scattering, the rate of electron scattering due to phonons in a semi-classical approximation is given by: \color{black} \cite{lundstrombook,book1}: 
\begin{eqnarray}
\frac{\partial f(r,\overrightarrow{k},t)}{\partial t}=\sum_{\overrightarrow{k'}}\Big\{\underbrace{S(\overrightarrow{k'},\overrightarrow{k})\{1-f(r,\overrightarrow{k},t)\}  f(r,\overrightarrow{k'},t)}_{in-scattering} \nonumber\\
 -\underbrace{S(\overrightarrow{k},\overrightarrow{k'})\{1-f(r,\overrightarrow{k'},t)\}f(r,\overrightarrow{k},t)}_{out-scattering}\Big\} \delta(E_k-E_{k'}) \nonumber \\ 
 \label{eq:boltzmann}
\end{eqnarray}
$S(\overrightarrow{k},\overrightarrow{k'})/S(\overrightarrow{k'},\overrightarrow{k})$ incorporate the dependence of the rate of electron scattering on energy/momentum. For isotropic scattering with acoustic phonons, $S(\overrightarrow{k'},\overrightarrow{k})$ is independent of $\overrightarrow{k'}$ or $\overrightarrow{k}$
\begin{eqnarray}
S(\overrightarrow{k'},\overrightarrow{k})=S(\overrightarrow{k},\overrightarrow{k'})=S(E_{\overrightarrow{k}})=S(E_{\overrightarrow{k'}})\nonumber \\
=\frac{2\pi k_BTD_{ac}^2}{\rho \hslash v_s^2A},
\label{eq:s}
\end{eqnarray}
where   $D_{ac}$, $\rho$ and $v_s$ are the acoustic deformation potential, the mass density and the velocity of sound in the medium respectively  \cite{lundstrombook,book1}. The right side of Eq. \ref{eq:boltzmann} can be simplified by summing over the states  (assuming steady state) \cite{lundstrombook,book1}:

\begin{eqnarray}
\frac{\partial f(r,\overrightarrow{k},t)}{\partial t}=\{1-f(r,\overrightarrow{k})\}\sum_{\overrightarrow{k'}}S(\overrightarrow{k'},\overrightarrow{k})  f(r,\overrightarrow{k'}) \delta(E_k-E_{k'}) \nonumber\\
 -f(r,\overrightarrow{k})\sum_{\overrightarrow{k'}}S(\overrightarrow{k},\overrightarrow{k'})\{1-f(r,\overrightarrow{k'})\} \delta(E_k-E_{k'}) \nonumber \\ 
 =\{1-f(r,\overrightarrow{k})\}S(E_{\overrightarrow{k}})\sum_{\overrightarrow{k'}}  f(r,\overrightarrow{k'}) \delta(E_k-E_{k'}) \nonumber\\
 -f(r,\overrightarrow{k})S(E_{\overrightarrow{k}})\sum_{\overrightarrow{k'}}\{1-f(r,\overrightarrow{k'})\} \delta(E_k-E_{k'}) \nonumber \\ 
\end{eqnarray}

\begin{eqnarray}
\Rightarrow \frac{\partial f(r,\overrightarrow{k})}{\partial t}
 =\{1-f(r,\overrightarrow{k})\}\underbrace{S(E_{\overrightarrow{k}})n_{tot}(r,E_{\overrightarrow{k}})}_{\frac{2\pi}{\hslash}\Sigma^{in}(E_{\overrightarrow{k}})} \nonumber \\
 -f(r,\overrightarrow{k})\underbrace{S(E_{\overrightarrow{k}})p_{tot}(r,E_{\overrightarrow{k}})}_{\frac{2 \pi}{\hslash}\Sigma^{out}(E_{\overrightarrow{k}})}, \nonumber \\ 
\end{eqnarray}
where 
\begin{gather}
n_{tot}(r,E_{\overrightarrow{k}})=\sum_{\overrightarrow{k'}}n(r,E_{\overrightarrow{k'}})\delta(E_{\overrightarrow{k}}-E_{\overrightarrow{k'}}) \nonumber \\
 p_{tot}(r,E_{\overrightarrow{k}})=\sum_{\overrightarrow{k'}}p(r,E_{\overrightarrow{k'}})\delta(E_{\overrightarrow{k}}-E_{\overrightarrow{k'}}) \nonumber 
\end{gather}
For acoustic phonon, $S(E_{\overrightarrow{k}})$ is independent of $E_{\overrightarrow{k}}$, $n_{tot}(E_{\overrightarrow{k}}) \approx D(E)f(r,E) $ and  $p_{tot}(E_{\overrightarrow{k}}) \approx D(E)\{1-f(r,E)\} $. Therefore, 
\begin{gather}
\tau(E_{\overrightarrow{k}}) \propto \frac{1}{\frac{\partial f(r,\overrightarrow{k})}{\partial t}} \propto \frac{1}{D(E)} \nonumber \\
\Rightarrow \tau(E_{\overrightarrow{k}}) \propto E_{\overrightarrow{k}}^n,
\end{gather}

where $n=0.5,~0,~-0.5$ for $1-D,~2-D$ and $3-D$ devices respectively. To demonstrate the effect of the   scattering which are of order higher than  phonon scattering, we choose 
\[
S(E_{\overrightarrow{k}})=kE_{\overrightarrow{k}}^{-u},
\]
where $k$ is a constant of proportionality  such that 
\[
\tau(E_{\overrightarrow{k}}) \propto E_{\overrightarrow{k}}^{n+u},
\]
where $n$ is same as defined above and $r=n+u$ being the order of the scattering process. In NEGF, we then use
\begin{gather}
\Sigma^{in}(r,E_{\overrightarrow{k}})=\frac{\hslash}{2\pi} S(E_{\overrightarrow{k}})\sum_{\overrightarrow{k'}}n(r,E_{\overrightarrow{k'}})\delta(E_{\overrightarrow{k}}-E_{\overrightarrow{k'}}) \nonumber \\
\Sigma^{out}(r,E_{\overrightarrow{k}})=\frac{\hslash}{2\pi} S(E_{\overrightarrow{k}})\sum_{\overrightarrow{k'}}p(r,E_{\overrightarrow{k'}})\delta(E_{\overrightarrow{k}}-E_{\overrightarrow{k'}}) \nonumber \\
\end{gather}

\section{Derivation of the factor $\Upsilon$}\label{appendix3}
In case of diffusive or incoherent transport without externally applied magnetic field, the dynamics of the electron system follows the quasi-distribution function given by \cite{ashcroft}

\begin{eqnarray}
f(\overrightarrow{k})=f_0(E_{\overrightarrow{k}})+\int^{\infty}_{0}P(\overrightarrow{k},\tau')\Big\{(-\frac{\partial f_0}{\partial E}) \overrightarrow{v}(\overrightarrow{k}).\Big(-e\overrightarrow{\mathlarger{\mathlarger{\mathlarger{\varepsilon}}}} \nonumber \\
- \nabla \mu -\frac{E-\mu}{T} \nabla T\Big)\Big\}d\tau', \nonumber \\
\label{eq:diffusive_f}
\end{eqnarray}
where $P(\overrightarrow{k},\tau')$ is the fraction of the electrons with wavevector $\overrightarrow{k}$ that donot suffer a scattering within the time period $\tau'$. For isotropic scattering, generally $P(\overrightarrow{k},\tau')$ takes the form \cite{ashcroft}:

\begin{equation}
P(\overrightarrow{k},\tau')=e^{\frac{-\tau'}{\tau(\overrightarrow{k})}}
\end{equation}
Generally for isotropic and local scattering processes, $\tau(\overrightarrow{k})$ depends on $\overrightarrow{k}$ through the energy $E_{\overrightarrow{k}}$. Therefore, 

\begin{equation}
P(\overrightarrow{k},\tau')=e^{{-\tau'}/{\tau(E_{\overrightarrow{k}})}}
\end{equation}
Equation \eqref{eq:diffusive_f} then becomes	
\begin{eqnarray}
f(\overrightarrow{k})=f_0(E_{\overrightarrow{k}})+\int^{\infty}_{0}e^{{-\tau'}/{\tau(E_{\overrightarrow{k}})}}\Big\{(-\frac{\partial f_0}{\partial E}) \overrightarrow{v}(\overrightarrow{k}).\Big(-e\overrightarrow{\mathlarger{\mathlarger{\mathlarger{\varepsilon}}}} \nonumber \\
- \nabla \mu -\frac{E-\mu}{T} \nabla T\Big)\Big\}d\tau' \nonumber \\
\Rightarrow f(\overrightarrow{k})=f_0(E_{\overrightarrow{k}})+\tau(E_{\overrightarrow{k}}) \overrightarrow{v}(\overrightarrow{k}).\Big\{(-\frac{\partial f_0}{\partial E})\Big(-e\overrightarrow{\mathlarger{\mathlarger{\mathlarger{\varepsilon}}}} 
- \nabla \mu \nonumber \\-\frac{E-\mu}{T} \nabla T\Big)\Big\} \nonumber \\
\end{eqnarray}
 \begin{figure}[!htb]
\includegraphics[scale=.26]{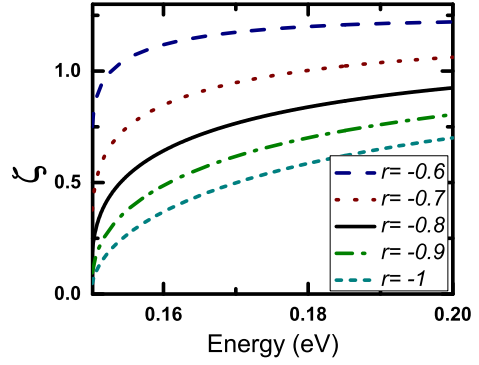}
\caption{Plot of the factor $\zeta=\frac{ (E^{\frac{3}{2}}-E_b^{\frac{3}{2}} )E^r}{ \{(E-E_b)^{r+\frac{3}{2}} \}}$ for various values of $r$ at $E_b=0.15eV$.}
\label{fig:zeta}
\end{figure}
Assuming that the potential and the temperature gradient are applied in the $z$ direction only,
\begin{eqnarray}
f(\overrightarrow{k})=f_0(E_{\overrightarrow{k}})+\tau(E_{\overrightarrow{k}}) {v_z}(\overrightarrow{k})\Big\{(-\frac{\partial f_0}{\partial E})\Big(-e\overrightarrow{\mathlarger{\mathlarger{\mathlarger{\varepsilon_z}}}} 
-\frac{\partial \mu(z)}{\partial z} \nonumber \\-\frac{E-\mu(z)}{T(z)} \frac{\partial T(z)}{\partial z} \Big)\Big\} \nonumber \\
\end{eqnarray}
The current density in the z-direction, is therefore, given by:
\[
{j_z}=-e\int \frac{d\overrightarrow{k}}{4 \pi^3}v_z(\overrightarrow{k})f(\overrightarrow{k}) 
\]
\begin{eqnarray}
=-e\int \frac{d\overrightarrow{k}}{4 \pi^3}v_z(\overrightarrow{k}) \Big[ f_0(E_{\overrightarrow{k}})+\tau(E_{\overrightarrow{k}}) {v_z}(\overrightarrow{k})\Big\{(-\frac{\partial f_0}{\partial E}) \nonumber \\ \Big(-e\overrightarrow{\mathlarger{\mathlarger{\mathlarger{\varepsilon_z}}}} 
-\frac{\partial \mu(z)}{\partial z}-\frac{E-\mu(z)}{T(z)} \frac{\partial T(z)}{\partial z} \Big)\Big\}\Big] \nonumber \\
\end{eqnarray}
The integral of the term $v_z(\overrightarrow{k}) f_0(E_{\overrightarrow{k}})$ vanishes since $f_0$ depends only on energy and is symmetrical in $\overrightarrow{k}$ space.

\begin{eqnarray}
j_z=-e\int \frac{d\overrightarrow{k}}{4 \pi^3} \tau(E_{\overrightarrow{k}}) \|\overrightarrow{v_z}(\overrightarrow{k})\|^2\Big\{(-\frac{\partial f_0}{\partial E})\Big(-e\overrightarrow{\mathlarger{\mathlarger{\mathlarger{\varepsilon_z}}}}  \nonumber \\
-\frac{\partial \mu(z)}{\partial z}-\frac{E-\mu(z)}{T(z)} \frac{\partial T(z)}{\partial z} \Big)\Big\} \nonumber \\
\label{eq:final_curr}
\end{eqnarray}
$\tau(E_{\overrightarrow{k}})$ and $\|\overrightarrow{v_z}(\overrightarrow{k})\|^2$ depend on $\overrightarrow{k}$ only through the energy $E_{\overrightarrow{k}}$. We can simplify \eqref{eq:final_curr} to transform $\overrightarrow{k}$ dependence to energy $(E)$ dependence:

\begin{eqnarray}
j_z=-e\int \tau(E) \|\overrightarrow{v_z}(E)\|^2 D(E)\Big\{(-\frac{\partial f_0}{\partial E})\Big(-e\overrightarrow{\mathlarger{\mathlarger{\mathlarger{\varepsilon_z}}}}  \nonumber \\
-\frac{\partial \mu(z)}{\partial z}-\frac{E-\mu(z)}{T(z)} \frac{\partial T(z)}{\partial z} \Big)\Big\}dE \nonumber \\
\end{eqnarray}

The term within the second bracket is the driving force for the current and the term $\tau(E) \|\overrightarrow{v_z}(E)\|^2 D(E)$ defines the ease with which the driving force can cause a flow of the current. An ideal energy filter should block the the current flow from the cold contact to the hot contact due to the negative part of the driving force  and therefore only filter out the current produced  due to the positive part of the driving force.  For the same applied voltage and temperature gradient and assuming an ideal energy filter, the overall current should be given by the equation:
\begin{eqnarray}
j_z=-e\int\limits_{\epsilon}^{\infty} \tau(E) \|\overrightarrow{v_z}(E)\|^2 D(E)\Big\{(-\frac{\partial f_0}{\partial E})\Big(-e\overrightarrow{\mathlarger{\mathlarger{\mathlarger{\varepsilon_z}}}}  \nonumber \\
-\frac{\partial \mu(z)}{\partial z}-\frac{E-\mu(z)}{T(z)} \frac{\partial T(z)}{\partial z} \Big)\Big\}dE, \nonumber \\
\end{eqnarray}
where $\epsilon$ is defined by the equation:
\[
\Big(-e\overrightarrow{\mathlarger{\mathlarger{\mathlarger{\varepsilon_z}}}}  \nonumber \\
-\frac{\partial \mu(z)}{\partial z}-\frac{\epsilon-\mu(z)}{T(z)} \frac{\partial T(z)}{\partial z} \Big)=0.
\]
 The generated power would increase with increase in the current. In case of an ideal filter,  a sufficient but not necessary condition for improvement of generated power with filtering is that $\tau(E) \|\overrightarrow{v_z}(E)\|^2D(E)$ is an increasing function of $E$. In other words,
\begin{equation}
\frac{\tau(E+E_b) \|\overrightarrow{v_z}(E+E_b)\|^2D(E+E_b)}{\tau(E) \|\overrightarrow{v_z}(E)\|^2D(E)}>1,
\label{eq:condition}
\end{equation}
for $E_b>0$. Here $E_b$ is the cut-off energy for filtering. For isotropic and local scattering processes, $\tau(E)$ can generally be approximated as $\tau(E)=\sum_i k_iE^{r_i}$. In case of single moded nanowires, $\|\overrightarrow{v_z}(E)\|^2D(E) =2\sqrt{\frac{2\pi E}{m_lh^2}} $. The minimum value of $r$ for which energy filtering can enhance the generated power  in case of perfect filtering is therefore $r>r_{min}=-\frac{1}{2}$. For imperfect filtering, the value of $r_{min}$ may further increase. \\
\indent For bulk generators, the value of $D(E)$ contributing to conduction cannot be defined properly due to partial momentum conservation. However assuming uncoupled mode transport, we can draw an upper limit on the value of  $r_{min}$. It can be shown that for perfect filtering and no effect of scattering near the barrier on the performance of the device, an assumption of uncoupled modes in electron transport  gives:
\begin{gather}
<v_z^2(E)D(E)>=\sum_m v_z^2(E-E_m)D_{1D}(E-E_m)  \nonumber \\
 = \int_0^{(E-E_b)} 2\frac{(E-E_m)}{m_l} \sqrt{\frac{2\pi m_l}{h^2}}\frac{1}{\sqrt{E-E_m}} \left(\frac{4\pi m_t}{h^2}dE_m\right) \nonumber \\
 =\frac{16\pi}{3}\frac{1}{h^3}\sqrt{\frac{2\pi m_t^2}{m_l}} (E^{\frac{3}{2}}-E_b^{\frac{3}{2}} ) ,\nonumber \\  
\end{gather}

for $E>E_b$. Here $<>$ denotes the average value of the argument and $m$ denotes all possible modes that are available for conduction.
Assuming $\tau(E)=k_oE^r$, \eqref{eq:condition} translates to:
\[
\zeta=\frac{ (E^{\frac{3}{2}}-E_b^{\frac{3}{2}} )E^r}{ \{(E-E_b)^{r+\frac{3}{2}} \}}>1
\]
It can be shown that for $E_b=0.15eV$, the above condition is valid for $r \gtrapprox -0.7$ (See Fig. \ref{fig:zeta}).

\end{document}